\documentclass[aps,twocolumn,showpacs]{revtex4}
\usepackage[dvips]{graphicx}
\newcommand{\sect}[1]{\setcounter{equation}{0}\section{#1}}

\begin{document}
\title{Statistical mechanics \\ in the context of
special relativity}
\author{G. Kaniadakis}
\email{kaniadakis@polito.it} \affiliation{Dipartimento di Fisica
and Istituto Nazionale di
 Fisica della Materia, \\ Politecnico di Torino,
Corso Duca degli Abruzzi 24, 10129 Torino, Italy}
\date{\today}

\begin {abstract} In ref. [Physica A {\bf 296}, 405 (2001)],
starting from the one parameter deformation of the exponential
function $\exp_{_{\{{\scriptstyle \kappa}\}}}(x)=
\left(\sqrt{1+\kappa^2x^2}+\kappa x\right)^{1/\kappa}$, a
statistical mechanics has been constructed which reduces to the
ordinary Boltzmann-Gibbs statistical mechanics as the deformation
parameter $\kappa$ approaches to zero. The distribution
$f=\exp_{_{\{{\scriptstyle \kappa}\}}}\!\big(\!\!-\beta
E+\beta\mu \big)$ obtained within this statistical mechanics
shows a power law tail and depends on the non specified parameter
$\beta$, containing all the information about the temperature of
the system. On the other hand, the entropic form $S_{\kappa}=\int
d^3 p \,(\,c_{\kappa}f^{1+\kappa}+c_{-\kappa}f^{1-\kappa})$, which
after maximization produces the distribution $f$ and reduces to
the standard Boltzmann-Shannon entropy $S_{0}$ as $\kappa
\rightarrow 0$, contains the coefficient $c_{\kappa}$ whose
expression involves, beside the Boltzmann constant, another non
specified parameter $\alpha$.

In the present effort we show that $S_{\kappa}$ is the unique
existing entropy obtained by a continuous deformation of $S_{0}$
and preserving unaltered its fundamental properties of concavity,
additivity and extensivity. These properties of $S_{\kappa}$
permit to determine unequivocally the values of the above
mentioned parameters $\beta$ and $\alpha$. Subsequently, we
explain the origin of the deformation mechanism introduced by
$\kappa$ and show that this deformation emerges naturally within
the Einstein special relativity. Furthermore, we extend the theory
in order to treat statistical systems in a time dependent and
relativistic context. Then, we show that it is possible to
determine in a self consistent scheme within the special
relativity the values of the free parameter $\kappa$ which
results to depend on the light speed $c$ and reduces to zero as
$c \rightarrow \infty$ recovering in this way the ordinary
statistical mechanics and thermodynamics. The novel statistical
mechanics here presented preserves unaltered the mathematical and
epistemological structure of the ordinary statistical mechanics
and is suitable to describe a very large class of experimentally
observed phenomena in low and high energy physics and in natural,
economic and social sciences. Finally, in order to test the
correctness and predictability of the theory, as working example
we consider the cosmic rays spectrum, which spans 13 decades in
energy and 33 decades in flux, finding a high quality agreement
between our predictions and observed data.

\end {abstract}


\maketitle

\sect{Introduction}

The following one-parameter deformations of the exponential and
logarithm functions
\begin{eqnarray}
\exp_{_{\{{\scriptstyle \kappa}\}}}(x)&=&
\left(\sqrt{1+\kappa^2x^2}+\kappa x\right)^{1/\kappa} \ ,
\label{I1} \\ \ln_{_{\{{\scriptstyle \kappa}\}}}(x)&=&
\frac{x^{\kappa}-x^{-\kappa}}{2\kappa} \ , \label{I2}
\end{eqnarray}
which reduce to the standard exponential and logarithm
respectively as the real deformation parameter $\kappa$
approaches to zero, has been introduced recently in Ref.
\cite{PHA01}. The above functions have many very interesting
properties \cite{PHA01,PLA01,PHA02,NAU} (some being identical to
the ones of the undeformed functions) that permit to construct a
statistical mechanics (and thermodynamics) which generalizes the
standard Boltzmann-Gibbs one. This $\kappa$-deformed statistical
mechanics preserves unaltered the structure of the ordinary one
and can be used to explain a very large class of experimentally
observed phenomena described by distribution functions exhibiting
power law tails. The areas where this formalism can be applied
include among others, low and high energy physics, astrophysics,
econophysics, geology, biology, mathematics, information theory,
linguistics, etc. \cite{ABED,EDI02,OLE1,OLE2,ECON}.

In ref. \cite{PHA01} it has been shown that the statistical
distribution
\begin{eqnarray}
f=\exp_{_{\{{\scriptstyle \kappa}\}}}(-\beta \,[\,E-\mu\,]\,) \ ,
\label{I3}
\end{eqnarray}
which generalizes the Boltzmann-Gibbs distribution, can be
obtained also by maximizing, after properly constrained, the
entropy
\begin{equation}
S_{\kappa}=\int d^n v
\,(\,c_{\kappa}f^{1+\kappa}+c_{-\kappa}f^{1-\kappa}) \ ,
\label{I4}
\end{equation}
which reduces to the standard entropy $S_{0}$ as the deformation
parameter approaches to zero. The coefficient $c_{\kappa}$, which
also absorbs the Boltzmann constant $k_{_{B}}$, depends on a free
parameter $\alpha$ (see Eq. (65) of ref. \cite{PHA01}) which
remains to be determined together with the parameter $\beta$ which
contains the information about the temperature $T$ of the system.

A first question which arises naturally is if it is possible and
how to find any criterion which allows us to fix the parameters
$\beta$ and $\alpha$  or at least express these in terms of the
deformation parameter $\kappa$, in order to reduce the free
parameters of the theory.

A second question regards the properties of the entropy
$S_{\kappa}$. It is well known that the Boltzmann-Shannon entropy
$S_{0}$ is concave, additive and extensive. We know that
$S_{\kappa}$ is concave with respect to the variable $f$, but
what happens about its additivity and extensivity? More in
general, beside the Boltzmann-Shannon entropy  other concave,
additive and extensive entropies do exist?

A third question is related to the physical mechanism which
originates the deformation introduced by the parameter $\kappa$.
In other words,  does a more fundamental theory exist where this
deformation emerges, or is it simply a purely mathematical tool?

A fourth question is if it is possible to extend the theory
originally developed in the framework of a classical kinetics to
treat statistical systems in the context of a relativistic
kinetics.

A fifth question regards the deformation parameter $\kappa$. This
parameter will continue to remain free or is it possible to
determine its value self consistently within the theory?

The present paper deals with the above questions and its purpose
is double. Firstly, we will show that $S_{\kappa}$ is the unique
existing, concave, additive and extensive entropy, beside the
Boltzmann-Shannon entropy. As we will see these properties of
$S_{\kappa}$ are sufficient to determine unequivocally the values
of parameters $\beta$ and $\alpha$. Secondly, we will show that
the deformation introduced by $\kappa$ is a purely relativistic
effect and then we will explain the deformation mechanism within
the Einstein special relativity. Then, we will formulate a
relativistic $\kappa$-kinetics and we will calculate the value of
$\kappa$.

Finally, in order to test the predictability and correctness of
theory here proposed  we will consider two sets of experimental
data. In particular we will analyze the cosmic rays spectrum which
spans 13 decades in energy and 33 decades in particle flux that,
as it is widely known, violates the Boltzmann-Gibbs statistics.
As we will see, we have a high quality agreement between the
theory and the observed data.

The paper is organized as it follows. In Sect. II,  generalizing
the approach proposed in ref.s \cite{PHA01,PLA01}, we introduce a
class of one parameter deformed structures and study their
mathematical properties.

In Sect. III, within this context, starting from the Jaynes
maximum entropy principle we consider the most general class of
deformed statistical mechanics preserving the main features of the
standard Boltzmann-Gibbs one.

In Sect. IV, we show that the entropy $S_{\kappa}$ introduced in
ref. \cite{PHA01} is the only one existing beside the
Boltzmann-Shannon entropy $S_{0}$ which is simultaneously
concave, additive and extensive. Then the statistical mechanics
and thermodynamics based on $S_{\kappa}$ can be viewed as a
natural extension of the Boltzmann-Gibbs one, recovering this
last as the deformation parameter $\kappa$ approaches to zero.

In Sect. V, we consider the mean properties of
$\kappa$-exponential and $\kappa$-logarithm which have a
fundamental role in the formulation of the new statistical
mechanics.

In Sect. VI, we extend the formalism to a time dependent and
relativistic context. In particular after introducing the
relativistic $\kappa$-kinetic evolution equation we study its
stationary state and prove the H-theorem.

In Sect. VII, we explain the origin of the $\kappa$-deformation
and show that it emerges naturally within the Einstein special
relativity.

In Sect. VIII, we propose an approach which permits to determine
within the special relativity the value of the parameter $\kappa$.

In Sect. IX, we compare two sets of experimental data with the
predictions of the present theory.

Finally in Sect. X, some concluding remarks are reported.

\sect{Deformed mathematics}

\subsection{Generator of the Deformation} Let $g(x)$ be an arbitrary
real function of the real variable $x$, that we call generator of
the deformation, having the following properties

\noindent i) $g(x)\in C^{\infty}({\bf R})$;

\noindent ii) $g(-x)=-g(x)$;

\noindent iii) $d g(x)/dx>0$;

\noindent iv) $g(\pm \infty)=\pm \infty$;

\noindent v) $g(x)\approx x$, for $x\rightarrow 0$.

Starting from the generator $g(x)$, we construct the real function
$x_{_{\{{\scriptstyle \kappa}\}}}$ of the real variable $x$ and
depending on the real parameter $\kappa$, as follows

\begin{eqnarray}
x_{_{\{{\scriptstyle \kappa}\}}}= \frac{1}{\kappa} \,{\rm
arcsinh}\, g(\kappa x)\ \ . \label{II1}
\end{eqnarray}
Its properties descent directly from the ones of the generator
$g(x)$:

\noindent i) $x_{_{\{{\scriptstyle \kappa}\}}}\in C^{\infty}({\bf
R})$;

\noindent ii) $(-x)_{_{\{{\scriptstyle
\kappa}\}}}=-x_{_{\{{\scriptstyle \kappa}\}}}$;

\noindent iii) $d x_{_{\{{\scriptstyle \kappa}\}}}/dx>0$;

\noindent iv) $(\pm \infty)_{_{\{{\scriptstyle \kappa}\}}}=\pm
\infty$;

\noindent v) $x_{_{\{{\scriptstyle \kappa}\}}}\approx x$, for
$x\rightarrow 0$ and then $0_{_{\{{\scriptstyle \kappa}\}}}=0$;

\noindent vi) $x_{_{\{{\scriptstyle \kappa}\}}}\approx x$, for
$\kappa \rightarrow 0$ and then $x_{_{\{{\scriptstyle 0}\}}}=x$;

\noindent vii) $x_{_{\{{\scriptstyle
-\kappa}\}}}=x_{_{\{{\scriptstyle \kappa}\}}}$.

Together with the function $x_{_{\{{\scriptstyle \kappa}\}}}$ one
can introduce the inverse function
$x^{\scriptscriptstyle{\{}{\scriptstyle\kappa}\scriptscriptstyle{\}}}$,
through $(x^{\scriptscriptstyle{\{}{\scriptstyle\kappa}
\scriptscriptstyle{\}}}) _{_{\{{\scriptstyle\kappa}\}}}=
(x_{_{\{{\scriptstyle\kappa}\}}})^{\scriptscriptstyle{\{}
{\scriptstyle\kappa}\scriptscriptstyle{\}}}=x$, which assumes the
form

\begin{eqnarray}
x^{\{{\scriptstyle \kappa}\}}= \frac{1}{\kappa}\, g^{-1}(\,\sinh
\, \kappa x)\ \ . \label{II2}
\end{eqnarray}

\subsection{Deformed algebra}

{\it Proposition 1:} The composition law
$\stackrel{\kappa}{\oplus}$ defined through
\begin{equation}
(x\stackrel{\kappa}{\oplus}y )_{_{\{{\scriptstyle
\kappa}\}}}=x_{_{\{{\scriptstyle \kappa}\}}}+y_{_{\{{\scriptstyle
\kappa}\}}}\ \ , \ \ \label{II3}
\end{equation}
which reduces to the ordinary sum as $\kappa\rightarrow 0$, namely
$x\stackrel{0}{\oplus}y=x+y$, is a deformed sum and the algebraic
structure $({\bf R},\stackrel{\kappa}{\oplus})$ forms an abelian
group.

{\it Proof:} Indeed from the definition of $x_{_{\{{\scriptstyle
\kappa}\}}}$ the following properties of
$\stackrel{\kappa}{\oplus}$ follow

\noindent 1) associativity property:
$(x\stackrel{\kappa}{\oplus}y)\stackrel{\kappa}{\oplus}z=x\stackrel{\kappa}{\oplus}(y\stackrel{\kappa}{\oplus}z)$;

\noindent 2) neutral element: $x\stackrel{\kappa}{\oplus}0=0
\stackrel{\kappa}{\oplus}x=x$;

\noindent 3) opposite element:
$x\stackrel{\kappa}{\oplus}(-x)=(-x)
\stackrel{\kappa}{\oplus}x=0$;

\noindent 4) commutativity property:
$x\stackrel{\kappa}{\oplus}y=y\stackrel{\kappa}{\oplus}x$.

Of course the $\kappa$-difference indicated with
$\stackrel{\kappa}{\ominus}$ is defined as
$x\stackrel{\kappa}{\ominus}y=x\stackrel{\kappa}{\oplus}(-y)$.

{\it Proposition 2:} The composition law
$\stackrel{\kappa}{\otimes}$ defined through
\begin{equation}
(x \otimes\!\!\!\!\!^{^{\scriptstyle \kappa}}\,\,y
)_{_{\{{\scriptstyle \kappa}\}}}=x_{_{\{{\scriptstyle
\kappa}\}}}\!\!\cdot y_{_{\{{\scriptstyle \kappa}\}}} \ \ ,
\label{II4}
\end{equation}
which reduces to the ordinary product as $\kappa\rightarrow 0$,
namely $x\stackrel{0}{\otimes}y=xy$, is a deformed product and
the algebraic structure $({\bf
R}-\{0\},\stackrel{\kappa}{\otimes})$ forms an abelian group.

{\it Proof:} Indeed from the definition of $x_{_{\{{\scriptstyle
\kappa}\}}}$ we have the following properties

\noindent 1) associative law: $(x \stackrel{\kappa}{\otimes}y)
 \stackrel{\kappa}{\otimes}z=x
 \stackrel{\kappa}{\otimes}(y
 \stackrel{\kappa}{\otimes}z)$;

\noindent 2) neutral element:
 $x \stackrel{\kappa}{\otimes}I=I
 \stackrel{\kappa}{\otimes}x= x$;

\noindent 3) inverse element:
  $x \stackrel{\kappa}{\otimes}\overline x=
\overline x \stackrel{\kappa}{\otimes}x=I$;

\noindent 4) commutative law: $x\stackrel{\kappa}{\otimes}y= y
 \stackrel{\kappa}{\otimes}x$;

\noindent being the neutral element $I=1^{\{\kappa\}}$ while the
inverse element of $x$ is $\overline x=(1/x_{_{\{{\scriptstyle
\kappa}\}}})^{\{\kappa\}}$.

Of course the $\kappa$-division $\stackrel{\kappa}{\oslash}$ is
defined as
$x\stackrel{\kappa}{\oslash}y=x\stackrel{\kappa}{\otimes}\overline
y$.

{\it Proposition 3:} The deformed sum $\stackrel{\kappa}{\oplus}$
and product $\stackrel{\kappa}{\otimes}$ obey the distributive law
\begin{equation}
z\stackrel{\kappa}{\otimes}(x \stackrel{\kappa}{\oplus}y) = (z
\stackrel{\kappa}{\otimes}x) \stackrel{\kappa}{\oplus}(z
\stackrel{\kappa}{\otimes}y) \ \ , \label{II5}
\end{equation}
and then the algebraic structure $({\bf R},\stackrel{\kappa}
{\oplus},\stackrel{\kappa}{\otimes})$ forms an abelian field.

{\it Proof:} This proposition follows from the (\ref{II3}) and
(\ref{II4}).

We remark that the field $({\bf
R},\stackrel{\kappa}{\oplus},\stackrel{\kappa}{\otimes})$ is
isomorphic with the field $({\bf R}, + ,\cdot)$. Moreover
$z\cdot(x \stackrel{\kappa}{\oplus}y) \neq (z\cdot x)
\stackrel{\kappa}{\oplus}(z\cdot y)$ and then the structure $({\bf
R},\stackrel{\kappa}{\oplus},\cdot)$ it is not a field.

{\it Proposition 4:} The function
$x^{\scriptscriptstyle{\{}{\scriptstyle\kappa}\scriptscriptstyle{\}}}$
has the two following properties
\begin{eqnarray}
&&x^{\scriptscriptstyle{\{}{\scriptstyle\kappa}\scriptscriptstyle{\}}}
\stackrel{\kappa}{\oplus}y
^{\scriptscriptstyle{\{}{\scriptstyle\kappa}\scriptscriptstyle{\}}}
=(\,x+y\,)^{\scriptscriptstyle{\{}
{\scriptstyle\kappa}\scriptscriptstyle{\}}} \ \ , \label{II6}
\\
&&x^{\scriptscriptstyle{\{}{\scriptstyle\kappa}\scriptscriptstyle{\}}}
\stackrel{\kappa}{\otimes}y
^{\scriptscriptstyle{\{}{\scriptstyle\kappa}\scriptscriptstyle{\}}}
=(\,x\,y\,)^{\scriptscriptstyle{\{}
{\scriptstyle\kappa}\scriptscriptstyle{\}}} \ \ . \label{II7}
\end{eqnarray}

{\it Proof:} These properties follow directly from the
definitions given by Eq.s (\ref{II3}) and (\ref{II4}).

{\it Proposition 5:} The function $x_{_{\{{\scriptstyle
\kappa}\}}}$ and its inverse
$x^{\scriptscriptstyle{\{}{\scriptstyle\kappa}\scriptscriptstyle{\}}}$
obey the following scaling laws:
\begin{eqnarray}
&&x'{_{{\scriptscriptstyle\{}{\scriptstyle
\kappa}'{\scriptscriptstyle \!\}}}}=z \, x_{_{\{{\scriptstyle
\kappa}\}}} \ \ , \label{II8}
\\
&&x'^{\scriptscriptstyle{\{}{\scriptstyle\kappa'}\scriptscriptstyle{\!\}}}
=z \,
x^{\scriptscriptstyle{\{}{\scriptstyle\kappa}\scriptscriptstyle{\}}}
\ \ , \label{II9}
\end{eqnarray}
with
\begin{eqnarray}
&&x'=z\, x  \ \ , \label{II10}\\
&&\kappa'=\kappa/z \ \ . \label{II11}
\end{eqnarray}

{\it Proof:} These laws follow from the definitions of the
functions $x_{_{\{{\scriptstyle \kappa}\}}}$ and
$x^{\scriptscriptstyle{\{}{\scriptstyle\kappa}\scriptscriptstyle{\}}}$.

{\it Proposition 6:} The pseudodistributive law
\begin{equation}
z\cdot(x \stackrel{\kappa}{\oplus}y)= (z\cdot x)
\stackrel{\,\,\kappa/z}{\oplus}(z\cdot y) \ \ , \label{II12}
\end{equation}
holds and then the structure $({\bf
R},\stackrel{\kappa}{\oplus},\cdot)$ is a pseudofield.

{\it Proof:} Indeed by using the propositions $1$ and $5$ one
obtains

\begin{eqnarray}
z\cdot(x \stackrel{\kappa}{\oplus}y)=&&z\cdot\left((x
\stackrel{\kappa}{\oplus}y)_{_{\{{\scriptstyle \kappa}\}}}\right)
^{\scriptscriptstyle{\{}{\scriptstyle\kappa}\scriptscriptstyle{\}}}
\nonumber \\ =&& z\cdot\left(x_{_{\{{\scriptstyle \kappa}\}}} +
y_{_{\{{\scriptstyle \kappa}\}}}\right)
^{\scriptscriptstyle{\{}{\scriptstyle\kappa}\scriptscriptstyle{\}}}
\nonumber \\ =&& z\cdot\left(\frac{1}{z}\,x'_{_{\{{\scriptstyle
\kappa'}\}}} + \frac{1}{z}\,y'_{_{\{{\scriptstyle
\kappa'}\}}}\right)
^{\scriptscriptstyle{\{}{\scriptstyle\kappa}\scriptscriptstyle{\}}}
\nonumber \\ =&& z\cdot\left(\frac{1}{z}\,(x'
\stackrel{\,\kappa'}{\oplus}y')_{_{\{{\scriptstyle
\kappa'}\}}}\right)
^{\scriptscriptstyle{\{}{\scriptstyle\kappa}\scriptscriptstyle{\}}}
\nonumber \\ =&&
 z\cdot\left(\left(\frac{1}{z}\,(x'
\stackrel{\,\kappa'}{\oplus}y')\right)_{_{\{{\scriptstyle
\kappa}\}}}\right)
^{\scriptscriptstyle{\{}{\scriptstyle\kappa}\scriptscriptstyle{\}}}
\nonumber
\\=&& x'\stackrel{\,\kappa'}{\oplus}y' \nonumber
\\ =&& (z \cdot x)\stackrel{\kappa /z}{\oplus}(z \cdot y)
\nonumber \ \ .
\end{eqnarray}

\subsection{\bf Deformed derivative}

Consider the two algebraic structures
$(X,\,\stackrel{\kappa}{\oplus},\,\cdot\,)$ and $(Y,\, +
,\,\cdot\,)$ with $X\equiv{\bf R}$ and $Y\equiv{\bf R}$. Let us
introduce the set of the functions ${\cal F}=\{f: X
\stackrel{f}\rightarrow Y \}$ with ${\cal F}\subseteq
C^{\infty}(X)$.

The $\kappa$-differential $d _{_{\{{\scriptstyle \kappa}\}}}x$ is
defined as
\begin{eqnarray}
d _{_{\{{\scriptstyle \kappa}\}}}x=\lim_{z\rightarrow x} x
\stackrel{\kappa}{\ominus}z \ . \label{II13}
\end{eqnarray}
and results to be
\begin{eqnarray}
 d _{_{\{{\scriptstyle \kappa}\}}}x=dx_{_{\{{\scriptstyle
\kappa}\}}} \ . \label{II14}
\end{eqnarray}

We define the $\kappa$-derivative for the functions of the set
${\cal F}$ through
\begin{equation}
\frac{d\,f(x)}{d_{_{\{{\scriptstyle \kappa}\}}}
x}=\lim_{z\rightarrow x}\frac{f (x)-f(z)}{\displaystyle{ x
\stackrel{\kappa}{\ominus}z}}\ \ , \label{II15}
\end{equation}
with $x,z\in X$ and $f(x),f(z)\in Y$.  We observe that the
${\kappa}$-derivative, which reduces to the usual one as the
deformation parameter ${\kappa}\rightarrow 0$, can be written in
the form
\begin{equation}
\frac{d \, f(x)}{d _{_{\{{\scriptstyle \kappa}\}}}x}=\frac{d \,
f(x)}{d \, x_{_{\{{\scriptstyle
\kappa}\}}}}=\frac{1}{d\,x_{_{\{{\scriptstyle
\kappa}\}}}/dx}\,\frac{d \, f(x)}{dx} \ , \label{II16}
\end{equation}
from which clearly it appears that the ${\kappa}$-derivative is
governed by the same rules of the ordinary one.

\subsection{Deformed exponential}

The $\kappa$-exponential $\exp_{_{\{{\scriptstyle
\kappa}\}}}(x)\in {\cal F}$ is defined as eigenstate of the
$\kappa$-derivative
\begin{equation}
\frac{d\,\exp_{_{\{{\scriptstyle \kappa}\}}}(x)}{d\,
x_{_{\{{\scriptstyle \kappa}\}}}}=\exp_{_{\{{\scriptstyle
\kappa}\}}}(x) \ \ , \label{II17}
\end{equation}
and is given by
\begin{eqnarray}
\exp_{_{\{{\scriptstyle \kappa}\}}}(x)=\exp
\,(x_{_{\{{\scriptstyle \kappa}\}}}) \ \ . \label{II18}
\end{eqnarray}

It results that
\begin{eqnarray}
&&\exp_{_{\{{\scriptstyle 0}\}}}(x)=\exp (x) \ \ , \label{II19}
\\
&&\exp_{_{\{{\scriptstyle - \kappa}\}}}(x)=\exp_{_{\{{\scriptstyle
\kappa}\}}}(x) \ \ . \label{II20}
\end{eqnarray}

The $\kappa$-exponential, just as the ordinary exponential, has
the properties
\begin{eqnarray}
&&\exp_{_{\{{\scriptstyle \kappa}\}}}(x) \in C^{\infty}({\bf R});
\label{II21}\\
&&\frac{d}{d\,x}\, \exp_{_{\{{\scriptstyle \kappa}\}}}(x)>0;
\label{II27}\\
&&\exp_{_{\{{\scriptstyle \kappa}\}}}(-\infty)=0^+;
\label{II22}\\
&&\exp_{_{\{{\scriptstyle \kappa}\}}}(0)=1;
\label{II24}\\
&&\exp_{_{\{{\scriptstyle \kappa}\}}}(+\infty)=+\infty;
\label{II26}\\
&&\exp_{_{\{{\scriptstyle \kappa}\}}}(x)\exp_{_{\{{\scriptstyle
\kappa}\}}}(-x)= 1 \ \ . \label{II28}
\end{eqnarray}

Furthermore, the $\kappa$-exponential has the two properties
\begin{eqnarray}
&&\left (\exp_{_{\{{\scriptstyle \kappa}\}}}(x)\right )^{r}
=\exp_{_{\{{\scriptstyle \kappa/r}\}}}(r x) \ \ , \label{II29}
\\
&&\exp_{_{\{{\scriptstyle \kappa}\}}}(x)\exp_{_{\{{\scriptstyle
\kappa}\}}}(y)=\exp_{_{\{{\scriptstyle
\kappa}\}}}(x\stackrel{\kappa}{\oplus}y)\ \ , \label{II30}
\end{eqnarray}
with $r\in {\bf R}$, and can be expressed in terms of the
generator $g(x)$ as
\begin{equation}
\exp_{_{\{{\scriptstyle \kappa}\}}}(x)= \left[\sqrt{1+g(\kappa
x)^{\,2}}+g(\kappa x)\right]^{1/\kappa}\ \ . \label{II31}
\end{equation}

\subsection{Deformed logarithm}

The $\kappa$-logarithm $\ln_{_{\{{\scriptstyle \kappa}\}}}(x)$ is
defined as the inverse function of the of $\kappa$-exponential,
namely  $\ln_{_{\{{\scriptstyle
\kappa}\}}}(\exp_{_{\{{\scriptstyle
\kappa}\}}}x)=\exp_{_{\{{\scriptstyle
\kappa}\}}}(\ln_{_{\{{\scriptstyle \kappa}\}}}x)=x$, and is given
by
\begin{eqnarray}
\ln_{_{\{{\scriptstyle \kappa}\}}}(x)= (\ln
x)^{\scriptscriptstyle{\{}{\scriptstyle\kappa}\scriptscriptstyle{\}}}
\ \ . \label{II32}
\end{eqnarray}

It results that
\begin{eqnarray}
&&\ln_{_{\{{\scriptstyle 0}\}}}(x)=\ln (x) \ \ , \label{II33}
\\
&&\ln_{_{\{{\scriptstyle - \kappa}\}}}(x)=\ln_{_{\{{\scriptstyle
\kappa}\}}}(x) \ \ . \label{II34}
\end{eqnarray}

The $\kappa$-logarithm, just as the ordinary logarithm, has the
properties
\begin{eqnarray}
&&\ln_{_{\{{\scriptstyle \kappa}\}}}(x) \in C^{\infty}({\bf R}^+);
\label{II35}\\
&&\frac{d}{d\,x}\, \ln_{_{\{{\scriptstyle \kappa}\}}}(x)>0;
\label{II41}\\
&&\ln_{_{\{{\scriptstyle \kappa}\}}}(0^+)=-\infty;
\label{II36}\\
&&\ln_{_{\{{\scriptstyle \kappa}\}}}(1)=0;
\label{II38}\\
&&\ln_{_{\{{\scriptstyle \kappa}\}}}(+\infty)=+\infty;
\label{II40}\\
&&\ln_{_{\{{\scriptstyle \kappa}\}}}(1/x)=-\ln_{_{\{{\scriptstyle
\kappa}\}}}(x) \ \ . \label{II42}
\end{eqnarray}

Furthermore, the $\kappa$-logarithm has the two properties
\begin{eqnarray}
&&\ln_{_{\{{\scriptstyle \kappa}\}}}(x^{r}) =r
\ln_{_{\{{\scriptstyle r \kappa}\}}}(x) \ \ , \label{II43}
\\
&&\ln_{_{\{{\scriptstyle \kappa}\}}}(x \,y)
=\ln_{_{\{{\scriptstyle \kappa}\}}}(x)
\oplus\!\!\!\!\!^{^{\scriptstyle
\kappa}}\,\,\ln_{_{\{{\scriptstyle \kappa}\}}}(y)\ \ ,
\label{II44}
\end{eqnarray}
with $r\in {\bf R}$, and can be expressed in terms of the
generator $g(x)$ as
\begin{equation}
\ln_{_{\{{\scriptstyle \kappa}\}}}(x)=\frac{1}{\kappa}
g^{-1}\left( \frac{x^{\kappa}-x^{-\kappa}}{2} \right) \ \ .
\label{II45}
\end{equation}

Eq. (\ref{II45}) defines a very large class of deformed logarithms
varying the arbitrary function $g(x)$. These deformed logarithms
can depend on many other parameter (through the generator $g(x)$)
besides the parameter $\kappa$. We recall briefly that in
literature one can find other one \cite{TLOG,ABE} or two
\cite{BORO} parameter deformations of the exponential and
logarithm functions. Anyway in following we will consider the
deformed logarithms defined through Eq. (\ref{II45}) and
depending only on the parameter $\kappa$.

\subsection{Deformed trigonometry}
We define the $\kappa$-hyperbolic sine and cosine
\begin{eqnarray}
\sinh_{_{\{{\scriptstyle \kappa}\}}}(x)
=\frac{1}{2}\left[\exp_{_{\{{\scriptstyle \kappa}\}}}(x)
-\exp_{_{\{{\scriptstyle \kappa}\}}}(-x)\right] \ \ , \label{II46} \\
\cosh_{_{\{{\scriptstyle \kappa}\}}}(x)
=\frac{1}{2}\left[\exp_{_{\{{\scriptstyle \kappa}\}}}(x)
+\exp_{_{\{{\scriptstyle \kappa}\}}}(-x)\right]  \ \ ,
\label{II47}
\end{eqnarray}
starting from the $\kappa$-Euler formula
\begin{eqnarray}
\exp_{_{\{{\scriptstyle \kappa}\}}}(\pm
x)=\cosh_{_{\{{\scriptstyle \kappa}\}}}(x)\pm
\sinh_{_{\{{\scriptstyle \kappa}\}}}(x) \ \ . \label{II48}
\end{eqnarray}
It is straightforward to introduce the $\kappa$-hyperbolic
trigonometry, which reduces to the ordinary one as $\kappa
\rightarrow 0$. For instance, the formulas
\begin{eqnarray}
&&\cosh_{_{\{{\scriptstyle \kappa}\}}}^2(x)-
\sinh_{_{\{{\scriptstyle \kappa}\}}}^2(x)=1 \ , \label{II49} \\
&&\tanh_{_{\{{\scriptstyle
\kappa}\}}}(x)=\frac{\sinh_{_{\{{\scriptstyle
\kappa}\}}}(x)}{\cosh_{_{\{{\scriptstyle \kappa}\}}}(x)} \ ,
\label{II50} \\
&&\coth_{_{\{{\scriptstyle
\kappa}\}}}(x)=\frac{\cosh_{_{\{{\scriptstyle
\kappa}\}}}(x)}{\sinh_{_{\{{\scriptstyle \kappa}\}}}(x)} \ ,
\label{II51}
\end{eqnarray}
still hold true. All the formulas of the ordinary hyperbolic
trigonometry still hold true after properly deformed. The
deformation of a given formula can be obtained starting from the
corresponding undeformed formula, and then by making in the
argument of the hyperbolic trigonometric functions the
substitutions $x+y\rightarrow x\stackrel{\kappa}{\oplus}y$, and
obviously $nx\rightarrow x\stackrel{\kappa}{\oplus}x ...
\stackrel{\kappa}{\oplus}x$ (n times). For instance it results
\begin{equation}
\sinh_{_{\{{\scriptstyle \kappa}\}}}(x\stackrel{\kappa}{\oplus}y
)\!+\!\sinh_{_{\{{\scriptstyle \kappa}\}}}(
x\stackrel{\kappa}{\ominus}y ) \!=\!2\sinh_{_{\{{\scriptstyle
\kappa}\}}}(x) \cosh_{_{\{{\scriptstyle \kappa}\}}}(y)\ ,
\end{equation}
\begin{equation} \tanh_{_{\{{\scriptstyle
\kappa}\}}}(x)+\tanh_{_{\{{\scriptstyle
\kappa}\}}}(y)=\frac{\sinh_{_{\{{\scriptstyle
\kappa}\}}}(x\stackrel{\kappa}{\oplus}y )}
{\cosh_{_{\{{\scriptstyle \kappa}\}}}(x)\cosh_{_{\{{\scriptstyle
\kappa}\}}}(y)} \ \ ,
\end{equation}
and so on.

The $\kappa$-De Moivre  formula involving hyperbolic trigonometric
functions having arguments of the type $r x$ with $r\in{\bf R}$,
assumes the form
\begin{equation}
[\cosh_{_{\{{\scriptstyle \kappa}\}}}\!(x)\!\pm\!
\sinh_{_{\{{\scriptstyle \kappa}\}}}\!(x)
]^{r}\!\!=\!\cosh_{_{\{{\scriptstyle \kappa}/{\scriptstyle
r}\}}}\!(r x)\!\pm\!\sinh_{_{\{{\scriptstyle \kappa}/{\scriptstyle
r}\}}}\!(r x) \ . \label{II52}
\end{equation}

Also the formulas involving the derivatives of the hyperbolic
trigonometric function still hold, after properly deformed. For
instance we have
\begin{eqnarray}
&&\frac{d \,\sinh_{_{\{{\scriptstyle \kappa}\}}}(x)
}{d\,x_{_{\{{\scriptstyle \kappa}\}}}}= \cosh_{_{\{{\scriptstyle
\kappa}\}}}(x) \ \ , \label{II53} \\
&&\frac{d \,\tanh_{_{\{{\scriptstyle \kappa}\}}}(x)
}{d\,x_{_{\{{\scriptstyle \kappa}\}}}}=
\frac{1}{[\,\cosh_{_{\{{\scriptstyle \kappa}\}}}(x)\,]^2}
\label{II54} \ \ ,
\end{eqnarray}
and so on.

The $\kappa$-cyclic trigonometry can be constructed analogously.
The $\kappa$-sine and $\kappa$-cosine is defined as
\begin{eqnarray}
&&\sin_{_{\{{\scriptstyle
\kappa}\}}}(x)=-i\sinh_{_{\{{\scriptstyle \kappa}\}}}(ix) \ , \\
&&\cos_{_{\{{\scriptstyle \kappa}\}}}(x)=\cosh_{_{\{{\scriptstyle
\kappa}\}}}(ix) \ .
\end{eqnarray}
We remark that it results: $\sin_{_{\{{\scriptstyle
\kappa}\}}}(x)=\sin (x_{_{\{{\scriptstyle i\kappa}\}}}) $ and
$\cos_{_{\{{\scriptstyle \kappa}\}}}(x)=\cos(x_{_{\{{\scriptstyle
i\kappa}\}}}) $.

\subsection{Deformed inverse functions}

The $\kappa$-inverse hyperbolic or cyclic trigonometric functions
can be introduced starting from the corresponding direct functions
just as in the case of the undeformed mathematics. It is trivial
to verify that $\kappa$-inverse functions are related to the
$\kappa$-logarithm by the usual formulas of standard mathematics.
For instance we have
\begin{eqnarray}
&&{\rm arcsin}_{_{\{{\scriptstyle
\kappa}\}}}(x)=-i\,\ln_{_{\{{\scriptstyle
\kappa}\}}}\left(ix+\sqrt{1-x^2}\right) , \label{II55} \\
&&{\rm arctanh}_{_{\{{\scriptstyle
\kappa}\}}}(x)=\frac{1}{2}\ln_{_{\{{\scriptstyle
\kappa}\}}}\frac{1+x}{1-x} \label{II56} \ \ ,
\end{eqnarray}
and so on.

\subsection{Deformed product and sum of functions}

Let us consider the set of the non negative real functions ${\cal
D}=\{f,h,w,...\}$.

{\it Proposition 7:} The composition law
$\otimes\mbox{\raisebox{-2mm}{\hspace{-2.5mm}$\scriptstyle
\kappa$}}\hspace{1mm}$ defined through
\begin{eqnarray}
f\otimes\mbox{\raisebox{-2mm}{\hspace{-3.3mm}$\scriptstyle
\kappa$}} \hspace{2mm}h= \exp_{_{\{{\scriptstyle
\kappa}\}}}\!\left(\,\ln_{_{\{{\scriptstyle
\kappa}\}}}f+\ln_{_{\{{\scriptstyle \kappa}\}}}h\right) \ \ ,
\label{II57}
\end{eqnarray}
which reduces to the ordinary product as $\kappa\rightarrow 0$,
namely
$f\otimes\mbox{\raisebox{-2.3mm}{\hspace{-2.7mm}$\scriptstyle 0$}}
\hspace{2mm}h= f\cdot h$, is a deformed product and the algebraic
structure $({\cal D}-\{0\},
\otimes\mbox{\raisebox{-2mm}{\hspace{-2.5mm}$\scriptstyle
\kappa$}}\hspace{1mm})$ forms an abelian group.

{\it Proof:} Indeed this product has the following properties
\\1) associative law:
$(f\otimes\mbox{\raisebox{-2mm}{\hspace{-3.3mm}$\scriptstyle
\kappa$}}
\hspace{2mm}h)\otimes\mbox{\raisebox{-2mm}{\hspace{-3.3mm}$\scriptstyle
\kappa$}}
\hspace{2mm}w=f\otimes\mbox{\raisebox{-2mm}{\hspace{-3.3mm}$\scriptstyle
\kappa$}} \hspace{2mm}
(h\otimes\mbox{\raisebox{-2mm}{\hspace{-3.3mm}$\scriptstyle
\kappa$}} \hspace{2mm}w)$;
\\ 2) neutral element:
$f\otimes\mbox{\raisebox{-2mm}{\hspace{-3.3mm}$\scriptstyle
\kappa$}}
\hspace{2mm}1=1\otimes\mbox{\raisebox{-2mm}{\hspace{-3.3mm}$\scriptstyle
\kappa$}} \hspace{2mm}f=f$;
\\ 3) inverse element:
$f\otimes\mbox{\raisebox{-2mm}{\hspace{-3.3mm}$\scriptstyle
\kappa$}} \hspace{2mm}(1/f)=
(1/f)\otimes\mbox{\raisebox{-2mm}{\hspace{-3.3mm}$\scriptstyle
\kappa$}} \hspace{2mm}f=1$;
\\ 4) commutative law: $f
\otimes\mbox{\raisebox{-2mm}{\hspace{-3.3mm}$\scriptstyle
\kappa$}}
\hspace{2mm}h=h\otimes\mbox{\raisebox{-2mm}{\hspace{-3.3mm}$\scriptstyle
\kappa$}} \hspace{2mm}f$.

Of course the division
$\oslash\mbox{\raisebox{-2mm}{\hspace{-2.2mm}$\scriptstyle
\kappa$}}\hspace{.5mm}$ can be defined through
$f\oslash\mbox{\raisebox{-2mm}{\hspace{-3.3mm}$\scriptstyle
\kappa$}}\hspace{2mm}h=f
\otimes\mbox{\raisebox{-2mm}{\hspace{-3.3mm}$\scriptstyle
\kappa$}} \hspace{1mm}\,(1/h)$. The deformed $\kappa$-power
$f^{\otimes r}$ is defined through
\begin{equation}
f^{\otimes r}=\exp_{_{\{{\scriptstyle
\kappa}\}}}\!\left(r\,\ln_{_{\{{\scriptstyle \kappa}\}}}f \right)
\ \ , \label{II58}
\end{equation}
and generalizes the ordinary power $f^r$. In particular, when $r$
is integer one has $f^{\otimes r}=
f\otimes\mbox{\raisebox{-2mm}{\hspace{-3.3mm}$\scriptstyle
\kappa$}} \hspace{2mm}f ...
\otimes\mbox{\raisebox{-2mm}{\hspace{-3.3mm}$\scriptstyle
\kappa$}} \hspace{2mm}f$, (r times).

{\it Proposition 8:} The algebraic structure $({\cal D},
\otimes\mbox{\raisebox{-2mm}{\hspace{-2.5mm}$\scriptstyle
\kappa$}}\hspace{1mm})$ forms an abelian monoid.

{\it Proof:} Indeed the element $0$ does not admit an inverse
element.

Furthermore, just as in the case of the ordinary product, it
results
$f\otimes\mbox{\raisebox{-2mm}{\hspace{-3.3mm}$\scriptstyle
\kappa$}}
\hspace{2mm}0=0\otimes\mbox{\raisebox{-2mm}{\hspace{-3.3mm}$\scriptstyle
\kappa$}} \hspace{2mm}f= 0$.

{\it Proposition 9:} The composition law
$\oplus\mbox{\raisebox{-2mm}{\hspace{-2.5mm}$\scriptstyle
\kappa$}} \hspace{1mm}$ defined through
\begin{equation}
f\oplus\mbox{\raisebox{-2mm}{\hspace{-3.3mm}$\scriptstyle
\kappa$}} \hspace{2mm}h=\exp_{_{\{{\scriptstyle
\kappa}\}}}\left\{\ln\Big[\exp\left(\ln_{_{\{{\scriptstyle
\kappa}\}}}f\right)+ \exp\left(\ln_{_{\{{\scriptstyle
\kappa}\}}}h\right)\Big]\right\} \ , \label{II59}
\end{equation}
which reduces to the ordinary sum as the deformation parameter
approaches to zero, namely $f
\oplus\mbox{\raisebox{-2.4mm}{\hspace{-3.3mm}$\scriptstyle0$}}
\hspace{2mm}h=f+h$, is a deformed sum and the algebraic structure
$({\cal
D},\otimes\mbox{\raisebox{-2mm}{\hspace{-2.5mm}$\scriptstyle
\kappa$}}\hspace{1mm})$ forms an abelian monoid.

{\it Proof:} Indeed this sum has the following properties
\\ 1) associative
law: $(f\oplus\mbox{\raisebox{-2mm}{\hspace{-3.3mm}$\scriptstyle
\kappa$}} \hspace{2mm}h)
\oplus\mbox{\raisebox{-2mm}{\hspace{-3.3mm}$\scriptstyle \kappa$}}
\hspace{2mm}w=f\oplus\mbox{\raisebox{-2mm}{\hspace{-3mm}$\scriptstyle
\kappa$}} \hspace{2mm}(h
\oplus\mbox{\raisebox{-2mm}{\hspace{-3.3mm}$\scriptstyle \kappa$}}
\hspace{2mm}w)$;\\ 2) neutral element:
$f\oplus\mbox{\raisebox{-2mm}{\hspace{-3.3mm}$\scriptstyle
\kappa$}} \hspace{2mm}0=0
\oplus\mbox{\raisebox{-2mm}{\hspace{-3.3mm}$\scriptstyle \kappa$}}
\hspace{2mm}f=f$;\\ 3) commutative law:
$f\oplus\mbox{\raisebox{-2mm}{\hspace{-3.3mm}$\scriptstyle
\kappa$}}
\hspace{2mm}h=h\oplus\mbox{\raisebox{-2mm}{\hspace{-3.3mm}$\scriptstyle
\kappa$}} \hspace{2mm}f$.

We remark that the product
$\otimes\mbox{\raisebox{-2mm}{\hspace{-2.5mm}$\scriptstyle
\kappa$}} \hspace{1mm}$ and sum
$\oplus\mbox{\raisebox{-2mm}{\hspace{-2.5mm}$\scriptstyle
\kappa$}} \hspace{1mm}$ are distributive operations
$w\otimes\mbox{\raisebox{-2mm}{\hspace{-3.3mm}$\scriptstyle
\kappa$}}
\hspace{2mm}(f\oplus\mbox{\raisebox{-2mm}{\hspace{-3.3mm}$\scriptstyle
\kappa$}} \hspace{2mm}h)=
(w\otimes\mbox{\raisebox{-2mm}{\hspace{-3.3mm}$\scriptstyle
\kappa$}} \hspace{2mm}f)
\oplus\mbox{\raisebox{-2mm}{\hspace{-3.3mm}$\scriptstyle \kappa$}}
\hspace{2mm}(w\otimes\mbox{\raisebox{-2mm}{\hspace{-3.3mm}$\scriptstyle
\kappa$}} \hspace{2mm}h)$.

The product
$\otimes\mbox{\raisebox{-2mm}{\hspace{-2.5mm}$\scriptstyle
\kappa$}} \hspace{1mm}$ allows us to write the following property
of the $\kappa$-exponential
\begin{eqnarray} \exp_{_{\{{\scriptstyle
\kappa}\}}}(x)\otimes\mbox{\raisebox{-2mm}{\hspace{-3.3mm}$\scriptstyle
\kappa$}} \hspace{2mm} \!\exp_{_{\{{\scriptstyle
\kappa}\}}}(y)=\exp_{_{\{{\scriptstyle \kappa}\}}}(x+y) \ .
\label{II60}
\end{eqnarray}

Equivalently Eq. (\ref{II60}) can be written also in the form
\begin{eqnarray}
\ln_{_{\{{\scriptstyle
\kappa}\}}}(f\otimes\mbox{\raisebox{-2mm}{\hspace{-3.3mm}$\scriptstyle
\kappa$}} \hspace{2mm}h)=\ln_{_{\{{\scriptstyle \kappa}\}}}(f)+
\ln_{_{\{{\scriptstyle \kappa}\}}}(h) \ . \label{II61}
\end{eqnarray}
Eq. (\ref{II61}) gives a relevant property for the
$\kappa$-logarithm.

Finally, starting from the definition of the $\kappa$-power
$f^{\otimes r}$, we obtain the following relation
\begin{eqnarray}
r\, \ln_{_{\{{\scriptstyle \kappa}\}}}(f)=\ln_{_{\{{\scriptstyle
\kappa}\}}}(f^{\otimes r}) \ \ . \label{II62}
\end{eqnarray}

The relations given by Eq.s (\ref{II61}) and (\ref{II62}), which
express two mathematical properties of the $\kappa$-logarithm,
will be very useful in the next section in defining a new
additive and extensive entropy.

\sect{The Jaynes Maximum Entropy Principle}

Let us consider the following non normalized statistical
distribution involving the $\kappa-$exponential

\begin{equation}
f=\exp_{_{\{{\scriptstyle \kappa}\}}}\!\big(\!\!-\beta(E-\mu)\big)
\ \ . \label{III1}
\end{equation}
We write the real non specified parameter $\beta$ as
\begin{equation}
\beta=\frac{1}{\lambda\,k_{_B}T} \ \ , \label{III2}
\end{equation}
being $\lambda$ a new real parameter, $k_{_B}$ the Boltzmann
constant and $T$ the temperature of the system.

In the following it will be useful to introduce the distribution
\begin{equation}
n=\alpha \,\exp_{_{\{{\scriptstyle
\kappa}\}}}\!\left(-\frac{E-\mu}{\lambda\,k_{_B}T}\right) \ \ ,
\label{III5}
\end{equation}
which is related with $f$ through $n=\alpha\,f$, being $\alpha$
another new real parameter which will be determined together with
$\lambda$ in the following.

We recall that in the ordinary statistical mechanics the mean
value of a given physical quantity $A(p,n)$ depending on the
variable $p$ and the distribution $n=n(p)$, is defined as
\begin{equation}
<A(p,n)\!>\,= \frac{\int d^3 p \, A(p,n) \, n(p)}{\int d^3 p  \,
\,n(p)} \ \ . \label{III7}
\end{equation}
Analogously, in the case where $A=A(p_1,p_2,n_1,n_2)$ depends on
two independent variables $p_1$, $p_2$ and on the two independent
distribution functions $n_1(p_1)$ and $n_2(p_2)$,  we have that
the mean value is given by
\begin{equation}
<A>\,= \frac{\int d^3 p_1\, d^3 p_2 \,\, A \, \, n_1(p_1) \,
n_2(p_2)}{\int d^3 p_1 \,d^3 p_2  \, n_1(p_1)\,n_2(p_2)}  \ \ .
\label{III8}
\end{equation}

It is easy to verify that the stationary distribution $n$ can be
obtained as solution of the following variational equation
\begin{equation}
\frac{\delta}{\delta n}\int d^3 p \left[-k_{_B}\,\lambda\int \!
\ln_{_{\{{\scriptstyle \kappa}\}}}(n/\alpha) \,\, dn -\frac{1}{T}
\,E\, n + \frac{\mu}{T} \, n \, \right ]=0\ \ . \label{III6}
\end{equation}
Then the distribution $n$ can be viewed as maximizing the
information content $I_{\kappa}$
\begin{equation}
I_{\kappa}=\int d^3 p \, \, J_{\kappa}(n) \ \ ; \ \
J_{\kappa}(n)= \lambda \int \!\ln_{_{\{{\scriptstyle
\kappa}\}}}(n/\alpha) \,\, dn \ , \label{III6a}
\end{equation}
under the constraints
\begin{equation}
\int d^3 p \,\,E \,\,n = U \ , \label{III6b}
\end{equation}
\begin{equation}
\int d^3 p \,\, n =N \ , \label{III6c}
\end{equation}
imposing the conservation of the mean energy and of the particle
number respectively. Note that the chemical potential $\mu$ should
be chosen in such a way to set the particle number equal to unity
namely $N=1$ \cite{NAU}.

We observe that when $\kappa=0$ it results $J_{0}(n)=n\ln\,n$ and
the information content $I_{0}$ is the mean value of the ordinary
logarithm. In this case the above variational equation expresses
the Jaynes maximum entropy principle which conducts to the
Boltzman-Gibbs statistical mechanics. In the following, in analogy
with the standard statistical mechanics, we require that
$I_{\kappa}$ must be expressed as the mean value of
$\ln_{_{\{{\scriptstyle \kappa}\}}}\!n$. To do so we must consider
the subclass of the deformed logarithms obeying the condition
\begin{equation}
\lambda \int \ln_{_{\{{\scriptstyle \kappa}\}}}(n/\alpha) \,\, dn
=n\, \ln_{_{\{{\scriptstyle \kappa}\}}}\!n \ \ . \label{III9}
\end{equation}
The above condition, as we will see in the following, allows the
simultaneous determination of both the form of $\kappa$-logarithm
and  the values of the free parameters $\alpha$ and $\lambda$.
Firstly we observe that this class contains the standard logarithm
$\ln n$, for which results $\kappa=0$, $\alpha=1/e$, $\lambda=1$.
It will be the task of the next section to investigate on the
existence of new additional solutions of Eq. (\ref{III9}), beside
the standard logarithm. Taking into account this condition we can
write the variational equation (\ref{III6}) in the form
\begin{equation}
\frac{\delta}{\delta n}\int d^3 p \left(-k_{_B}\, n\,
\ln_{_{\{{\scriptstyle \kappa}\}}}\!n  -\frac{1}{T} \,E \,n +
\frac{\mu}{T} \, n \, \right )=0\ \ . \label{III10}
\end{equation}
We define the $\kappa$-entropy through
\begin{equation}
S_{\kappa}= -k_{_B}\int d^3 p\,\,\, n\, \ln_{_{\{{\scriptstyle
\kappa}\}}}\!n \ \ , \label{III12}
\end{equation}
so that $S_{\kappa}$ can be viewed as proportional to the mean
value of the $\ln_{_{\{{\scriptstyle \kappa}\}}}\!n$, namely
\begin{equation}
S_{\kappa}= -\,k_{_B}\!<\ln_{_{\{{\scriptstyle \kappa}\}}}\!n \!>
\ \ . \label{III13}
\end{equation}
In this definition of $S_{\kappa}$ we have a perfect analogy with
the Shannon entropy $S_{0}$ which is the proportional to the mean
value of the $\ln \!n$. It is remarkable that in both the
definitions of $S_{\kappa}$ and $S_{0}$ appears the standard mean
value given by Eq. (\ref{III7}).

Eq. (\ref{III10}) assumes the form
\begin{equation}
\frac{\delta}{\delta n}\left(-\,k_{_B}\!<\ln_{_{\{{\scriptstyle
\kappa}\}}}\!n \!> -\frac{1}{T} \,<E>+ \frac{\mu}{T} \, \right
)=0 \ , \label{III11}
\end{equation}
and then
\begin{equation}
\frac{\delta}{\delta n}\left( S_{\kappa} -\frac{1}{T} \,U+
\frac{\mu}{T} \, \right )=0 \ . \label{III11}
\end{equation}
The above variational equation can be viewed as defining a
maximum entropy principle analogous of the Jaynes principle of
the standard Boltzmann-Gibbs statistical mechanics \cite{JAY1}. We
remark that this maximum entropy principle, in the form given by
Eq. (\ref{III10}), holds only and exclusively for the subclass of
$\kappa$-logarithms, which are solutions of the integral equation
(\ref{III9}).

We show now that the families of entropies, defined through
(\ref{III12}) and involving the $\kappa$-logarithms which are
solutions of Eq. (\ref{III9}), have two important properties
typical of the Shannon entropy. To do so we consider the
properties (\ref{II61}) and (\ref{II62}) of $\kappa$-logarithm,
which rearrange as
\begin{eqnarray}
&&\ln_{_{\{{\scriptstyle \kappa}\}}}\!n_1+ \ln_{_{\{{\scriptstyle
\kappa}\}}}\!n_2=
\ln_{_{\{{\scriptstyle \kappa}\}}}\!n_{12} \ \ , \label{III14} \\
&&r\,\ln_{_{\{{\scriptstyle \kappa}\}}}\!n=\ln_{_{\{{\scriptstyle
\kappa}\}}}\!n^{*} \ \ , \label{III15}
\end{eqnarray}
with
$n_{12}=n_1\otimes\mbox{\raisebox{-2mm}{\hspace{-3.9mm}$\scriptstyle
\,\kappa$}} \hspace{2mm}\!n_2$ and $n^*=n^{\otimes r}$. When the
systems $1$ and $2$ described through $n_1$ and $n_2$
respectively are statistically independent, and after taking into
account the definitions of the mean values (\ref{III7}),
(\ref{III8}) and of the $\kappa$-entropy (\ref{III13}), the two
above properties of $\kappa$-logarithm transform into the
following properties for the $\kappa$-entropy
\begin{eqnarray}
&&S_{\kappa}[n_1]+ S_{\kappa}[n_2]=
S_{\kappa}[n_{12}] \ \ , \label{III16} \\
&&r\,S_{\kappa}[n]= S_{\kappa}[n^{*}] \ \ \label{III17} .
\end{eqnarray}
Eq.s (\ref{III16}) and (\ref{III17}) say that the entropies
$S_{\kappa}$ defined starting from the $\kappa$-logarithms which
are solutions of Eq. (\ref{III9}) are additive and extensive just
as the Shannon entropy. The distribution $n_{12}$ describes the
composite system obtained starting from the systems $1$ and $2$
while $n^*$ describes the scaled system related to the system
described through $n$. Note that the state described through the
distribution $n_{12}$ is different with respect to the state
described through the distribution $n_1n_2$ resulting $S_{\kappa}
[n_{12}]\leq S_{\kappa} [n_1 n_2]$.

Finally, from the concavity property of the deformed
$\kappa$-logarithm  the concavity of $S_{\kappa}$, follows
\begin{eqnarray}
S_{\kappa} [\,tn_1+(1-t)n_2\,]\geq t\,S_{\kappa} [n_1]+ (1-t)
S_{\kappa} [n_2] \ \ , \label{III18}
\end{eqnarray}
with $0\leq t\leq 1$.

As we have already noted, the ordinary logarithm is solution of
Eq. (\ref{III9}) and then the Shannon entropy
\begin{equation}
S_{0}[n]=-k_{_B}\int d^3 p \, \,\, n\,\ln n   \ \ , \label{III19}
\end{equation}
which is additive and extensive ($n_{12}=n_1n_2$ and $n^{*}=n^r$),
is admitted within the present formalism.

In the next section we will show that Eq. (\ref{III9}) admits a
new (only one) less evident solution.  Then beside the Shannon
entropy we have a new concave, additive and extensive entropy
which is the $\kappa$-entropy proposed in ref. \cite{PHA01}.

\sect{The New Additive and Extensive Entropy}

We consider Eq. (\ref{III9}) which, after performing a derivation
with respect to $n$, assumes the form
\begin{equation}
n\frac{d}{d\,n}\ln_{_{\{{\scriptstyle
\kappa}\}}}\!n+\ln_{_{\{{\scriptstyle \kappa}\}}}\!n-\lambda
\ln_{_{\{{\scriptstyle \kappa}\}}}\!(n/\alpha)=0 \ \ . \label{IV1}
\end{equation}
In the following we will determine the explicit form of
$\ln_{_{\{{\scriptstyle \kappa}\}}}\!n$ by solving this
differential-functional equation. We recall that
$\ln_{_{\{{\scriptstyle \kappa}\}}}\!n$ can be expressed in terms
of the generator function according to Eq. (\ref{II45}) so that
Eq. (\ref{IV1}) becomes
\begin{eqnarray}
n\frac{d}{d\,n} \,\,g^{-1}\left( \frac{n^{\kappa}-n^{-\kappa}}{2}
\right)+ g^{-1}\left( \frac{n^{\kappa}-n^{-\kappa}}{2}
\right)&&\nonumber \\
-\lambda\,\, g^{-1}\left(
\frac{(n/\alpha)^{\kappa}-(n/\alpha)^{-\kappa}}{2} \right)=0&& .
\label{IV2}
\end{eqnarray}
In the above equation the function to be determined is now the
generator function $g$. To do so we make the following changes of
variables
\begin{eqnarray}
&& t=\kappa \ln n \ \ , \label{IV3} \\ && z(t)=g^{-1}(\sinh t) \ \
, \label{IV4} \\ && c=-\kappa \ln \alpha \ \ , \label{IV5}
\end{eqnarray}
so that Eq. (\ref{IV2}) assumes the following simple form
\begin{eqnarray}
\kappa \, z'(t) +  z(t) - \lambda\,z(t+c) =0 \ \ .\label{IV6}
\end{eqnarray}

The property (v) of the generator $g(x)$ imposes that $z(t)$
obeys the two conditions $z(0)=0$ and $z'(0)=1$.  These
conditions, if combined with Eq. (\ref{IV6}), can be equivalently
written in the form
\begin{eqnarray}
&& \lambda\,z(c)=\kappa \ \ , \label{IV7} \\ &&\lambda\,z'(c)=1 \
\ . \label{IV8}
\end{eqnarray}

It is more convenient to take into account these conditions and
write Eq. (\ref{IV6}) under the form
\begin{eqnarray}
z(t+c)=z(t)\,z'(c)+z'(t)\,z(c) \ \ . \label{IV9}
\end{eqnarray}
After recalling the property (ii) of $g(x)$, which imposes that
$z(-t)=-z(t)$, Eq. (\ref{IV9}) can be written as
\begin{eqnarray}
2\,z(c)\,z'(t)=z(c+t)+z(c-t) \ \ . \label{IV10}
\end{eqnarray}
Let us introduce the new function $w(t)=z'(t)$. We can see that
Eq. (\ref{IV10}), after deriving with respect to $c$, transforms
into the following functional equation
\begin{eqnarray}
2\,w(c)\,w(t)=w(c+t)+w(c-t) \ \ . \label{IV11}
\end{eqnarray}
Finally the nonlinear transformation defined through $w(t)=\cosh
\xi(t)$ permits us to write the last equation under the form
\begin{eqnarray}
&&\cosh \big[\xi(c)+\xi(t)\big]+\cosh \big[\xi(c)-\xi(t)\big]
\nonumber \\ &&=\cosh \big[\xi(c+t)\big]+\cosh \big[\xi(c-t)\big]
\ \ . \label{IV12}
\end{eqnarray}
It is trivial to verify that the most general solution of Eq.
(\ref{IV12}) is given by
\begin{eqnarray}
\xi(t)=r\,t \ \ , \label{IV13}
\end{eqnarray}
with $r$ an arbitrary real parameter.

{\it Shannon solution}: We note that in the case $r=0$ we obtain
$g(x)=\sinh x$ and then $\ln_{_{\{{\scriptstyle
\kappa}\}}}(n)=\ln n$. This is the well known standard logarithm
which, inserted in Eq. (\ref{III12}), produces the Shannon
entropy.

{\it The new solution}: We consider now the case $r\neq 0$. It is
easy to realize that in this case the generator $g(x)$ assumes
the form
\begin{eqnarray}
g(x)=\sinh\left[\frac{1}{r}\,\,{\rm arcsinh}\,(rx)\right] \ \ .
\label{IV14}
\end{eqnarray}
For simplicity of the exposition we firstly discuss the case
$r=1$ for which $g(x)=x$. After some simple calculations we obtain
that only when
\begin{eqnarray}
&&-1<\kappa<1 \ \ , \label{IV15}
\end{eqnarray}
it is possible to determine the real constants $\lambda$ and
$\alpha$ obtaining
\begin{eqnarray}
&&\lambda= \sqrt{1-\kappa^2}\ \ , \label{IV16} \\
&&\alpha=\left(\frac{1-\kappa}{1+\kappa}\right)^{1/2\kappa}
 \ \ . \label{IV17}
\end{eqnarray}
In this case the generator $g(x)=x$ imposes the following
expressions for the deformed logarithm and exponential
\begin{eqnarray}
&&\ln_{_{\{{\scriptstyle \kappa}\}}}(x)=
\frac{x^{\kappa}-x^{-\kappa}}{2\kappa} \ \ , \label{IV18} \\
&&\exp_{_{\{{\scriptstyle \kappa}\}}}(x)=
\left(\sqrt{1+\kappa^2x^2}+\kappa x\right)^{1/\kappa} \ \ .
\label{IV19}
\end{eqnarray}
For the general case where $r\neq 1$ we obtain the same solution
given by Eq.s (\ref{IV15})-(\ref{IV19}) with the only difference
that in place of $\kappa$  now the scaled parameter $r\kappa$
appears. Then we can set $r=1$ without loosing the generality of
the theory. Moreover for $\kappa=0$ we obtain the Shannon
solution as a particular and limiting case of the new solution.

Finally, after inserting the expression of the $\kappa$-logarithm
given by Eq. (\ref{IV18}) into Eq. (\ref{III12}), we can write the
new additive and extensive entropy in the following simple form
\begin{eqnarray}
S_{\kappa} [n]=-\,k_{_B}\int
d^3p\,\,\,\frac{n^{1+\kappa}-n^{1-\kappa}}{2\kappa} \ \ .
\label{IV20}
\end{eqnarray}

We can write the entropy $S_{\kappa}$ also in terms of the
distribution $f$ obtaining
\begin{eqnarray}
S_{\kappa} [f]=-k_{_B}\int d^3p\,\,\,\left(
c_{\kappa}f^{1+\kappa} + c_{-\kappa}f^{1-\kappa}\right) \ \
\label{IV21}
\end{eqnarray}
where the coefficient $c_{\kappa}=\alpha^{1+\kappa}/2\kappa$
depends exclusively on the deformation parameter $\kappa$ and is
given by
\begin{eqnarray}
c_{\kappa}=\frac{1}{2\kappa}
\left(\frac{1-\kappa}{1+\kappa}\right)^{\frac{1+\kappa}{2\kappa}}
 \ \ . \label{IV22}
\end{eqnarray}
The above entropy is contained, as a particular case, in the class
of entropies introduced previously in ref. \cite{PHA01} (In Eq.
(65) of this reference it appears the non specified parameter
$\alpha$, while the Boltzmmann constant is absent because setted
to have $k_{_B}(1+\kappa)\,\alpha=1$).

We recall that the entropy given by Eq. (\ref{IV20}) is different
from the nonextensive entropy introduced in ref.s
\cite{HACHA,NEXT}. Of course the statistical distribution defined
through Eq.s (\ref{III1}) and (\ref{IV19}), introduced previously
in ref. \cite{PHA01}, is also different from the distribution of
the nonextensive statistics \cite{NEXT} and of the plasmas physics
\cite{VASYL}.

\sect{The $\kappa$-exponential and $\kappa$-logarithm}

Let us report here the main mathematical properties, some of these
reported in ref.s \cite{PHA01,PLA01,PHA02}, of the functions
$\exp_{_{\{{\scriptstyle \kappa}\}}}(x)$ and
$\ln_{_{\{{\scriptstyle \kappa}\}}}(x)$ defined through Eq.s
(\ref{IV18}) and (\ref{IV19}), respectively.

We start by observing that the generator of the deformation is
the function $g(x)=x$ and then from (\ref{II1}) and (\ref{II2}) we
obtain
\begin{eqnarray}
&&x_{_{\{{\scriptstyle \kappa}\}}}= \frac{1}{\kappa} \,{\rm
arcsinh}\,(\kappa x) \ \ , \label{V1} \\
&&x^{\scriptscriptstyle{\{}{\scriptstyle\kappa}\scriptscriptstyle{\}}}=
\frac{1}{\kappa} \,\sinh \,(\kappa x)\ \ . \label{V2}
\end{eqnarray}
We have also (\cite{MATHG}, p. 58)
\begin{eqnarray}
x_{_{\{{\scriptstyle \kappa}\}}}=x \,\,
F\left(\frac{1}{2},\frac{1}{2};\frac{3}{2};-\kappa^2 x^2\right) \
\ ; \ \ \kappa^2x^2\leq 1\ \ . \label{V3}
\end{eqnarray}

The following function
\begin{eqnarray}
[\,x\,]=\frac{q^x-q^{-x}} {q-q^{-1}} \ \ , \label{V4}
\end{eqnarray}
which has a central role in quantum group theory \cite{TUS,UBR},
results to be proportional to
$x^{\scriptscriptstyle{\{}{\scriptstyle\kappa}
\scriptscriptstyle{\}}}$, namely we have the relation
\begin{eqnarray}
[\,x\,]=\frac{1} {\ln_{_{\{{\scriptstyle \kappa}\}}}\!(e)}
\,\,x^{\scriptscriptstyle{\{}{\scriptstyle\kappa}\scriptscriptstyle{\}}}
 \ \ , \label{V5}
\end{eqnarray}
which can be written also in the form
\begin{eqnarray}
[\,x\,]=\frac{\ln_{_{\{{\scriptstyle \kappa}\}}}\!(e^x)}
{\ln_{_{\{{\scriptstyle \kappa}\}}}\!(e)} \ \ , \label{V6}
\end{eqnarray}
with $q=e^{\kappa}$. We note that the well known symmetry of
quantum group theory $q \leftrightarrow q^{-1}$ is related with
the symmetry $\kappa \leftrightarrow \! -\kappa$ of the present
theory. We also observe that, exploiting Eq.s (\ref{II6}) and
(\ref{II7}), we can obtain the two following properties of the
function $[\,x\,]$
\begin{eqnarray}
&&[\,x+y\,]=[\,x\,]\stackrel{\kappa '}{\oplus}[\,y\,] \ \ ,
\label{V7}\\
&&[\,x\,y\,]=[\,x\,]\stackrel{\kappa '}{\otimes}[\,y\,] \ \ ,
\label{V8}
\end{eqnarray}
with $\kappa '=(q-q^{-1})/2$.

The definitions of $\kappa$-sum and $\kappa$-product given through
(\ref{II3}) and (\ref{II4}) respectively transform as follows
\begin{equation}
x\stackrel{\kappa}{\oplus}y={1\over\kappa}\,\sinh\Big(\,{\rm
arcsinh}\,(\kappa x)+{\rm arcsinh}\,(\kappa y)\,\Big) \ ,
\label{V9}
\end{equation}
\begin{equation}
x\stackrel{\kappa}{\otimes}y={1\over\kappa}\,\sinh
\left(\,{1\over\kappa}\,\,{\rm arcsinh}\,(\kappa x)\,\,{\rm
arcsinh}\,(\kappa y)\,\right) \ . \label{V10}
\end{equation}
In particular the $\kappa$-sum assumes a very simple form
\begin{equation}
x \oplus\!\!\!\!\!^{^{\scriptstyle
\kappa}}\,\,y=x\sqrt{1+\kappa^2y^2}+y\sqrt{1+\kappa^2x^2} \ \ .
\label{V11}
\end{equation}

Starting directly from the $\kappa$-sum given by (\ref{V11}), one
obtains the following expressions for the $\kappa$-differential
and $\kappa$-derivative
\begin{eqnarray}
&&d x_{_{\{{\scriptstyle \kappa}\}}}=
\frac{d\,x}{\displaystyle{\sqrt{1+\kappa^2\,x^2} }} \ ,
\label{V12} \\
&&\frac{d \, f(x)}{d x_{_{\{{\scriptstyle
\kappa}\}}}}=\sqrt{1+\kappa^2\,x^2}\,\,\,\frac{d \, f(x)}{d \, x}
\ . \label{V13}
\end{eqnarray}

We consider now the functions $\exp_{_{\{{\scriptstyle
\kappa}\}}}(x)$ and $\ln_{_{\{{\scriptstyle \kappa}\}}}(x)$ which
can be written also as
\begin{eqnarray}
&&\exp_{_{\{{\scriptstyle \kappa}\}}}x\,= \exp\left(
\frac{1}{\kappa} \,{\rm arcsinh}\, \kappa x \right) \ \ , \label{V14} \\
&&\ln_{_{\{{\scriptstyle \kappa}\}}}(x) = \frac{1}{\kappa}\,\sinh
\, (\kappa \ln x)\ \ . \label{V15}
\end{eqnarray}
We remark the following concavity properties
\begin{eqnarray}
&&\frac{d^2}{d\,x^2}\, \exp_{_{\{{\scriptstyle \kappa}\}}}(x)>0 \
\ ; \ \
x \in {\bf R} \ \ , \label{V16} \\
&&\frac{d^2}{d\,x^2}\, \ln_{_{\{{\scriptstyle \kappa}\}}}(x)<0 \
\ ; \ \ x>0 \ \ . \label{V17}
\end{eqnarray}

A very interesting property of these functions is their power law
asymptotic behavior
\begin{eqnarray}
&&\exp_{_{\{{\scriptstyle \kappa}\}}}(x)
{\atop\stackrel{\textstyle\sim}{\scriptstyle x\rightarrow
\pm\infty}}\big|\,2\kappa x\big|^{\pm1/|\kappa|} \ \ , \label{V18}\\
&&\ln_{_{\{{\scriptstyle \kappa}\}}}(x)
{\atop\stackrel{\textstyle\sim}{\scriptstyle x\rightarrow
0^+}}-\frac{1}{2\,|\kappa|}\,x^{-|\kappa|} \ \ , \label{V19} \\
&&\ln_{_{\{{\scriptstyle \kappa}\}}}(x)
{\atop\stackrel{\textstyle\sim}{\scriptstyle x\rightarrow
+\infty}}\,\,\frac{1}{2\,|\kappa|}\,x^{|\kappa|} \ \ . \label{V20}
\end{eqnarray}

The Taylor expansion of $\kappa$-exponential is given by
\begin{equation}
\exp_{_{\{{\scriptstyle \kappa}\}}}(x) = \sum_{n=0}^{\infty}
a_n(\kappa)\,\frac{x^n}{n!} \ \ \ ; \ \ \ \kappa^2 x^2 < 1 \ \ ,
\label{V21}
\end{equation}
(\cite{MATHG}, p. 26), with the coefficients $a_n$ defined as
\begin{eqnarray}
&&a_0(\kappa)= 1 \ \ ; \ \ a_1(\kappa)= 1 \ \
, \ \  \nonumber \\
&&a_{2m}(\kappa)= \prod_{j=0}^{m-1}\Big(1-(2j)^2\kappa^2\Big) \ \
, \label{V22} \\ &&a_{2m+1}(\kappa)=
\prod_{j=1}^{m}\Big(1-(2j-1)^2\kappa^2\Big) \ \ . \nonumber
\end{eqnarray}
It results that $a_n(0)=1$ and $a_n(-\kappa)=a_n(\kappa)$. We
note that the first three terms in the above Taylor expansion are
the same  as the ordinary exponential, namely
\begin{equation}
\exp_{_{\{{\scriptstyle \kappa}\}}}(x) = 1+ x + \frac{x^2}{2} +
(1-\kappa^2)\,\frac{x^3}{3!}+...  \ \ . \label{V23}
\end{equation}

In fig.1,  the function $\exp_{_{\{{\scriptstyle \kappa}\}}}(-x)$
for a fixed value of $\kappa$ is plotted. We note that the bulk
of this function is very close to the standard exponential.
Indeed the Taylor expansion of $\exp_{_{\{{\scriptstyle
\kappa}\}}}(-x)$ is the same, up to second order, of the one of
$\exp(-x)$. The tail of $\exp_{_{\{{\scriptstyle \kappa}\}}}(-x)$
behaves as a power law. Between the bulk and the tail  an
intermediate region whose extension depends on the value of
$\kappa$ exists.
\begin{figure}[h]
\centerline{
\includegraphics[width=.8\columnwidth,angle=-90]{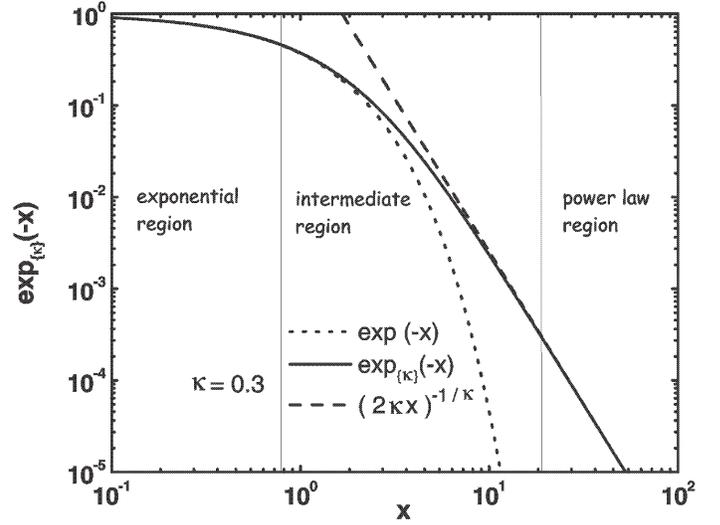}}
\caption{Plot of  the function $\exp_{_{\{{\scriptstyle
\kappa}\}}}(-x)$ versus $x$ for $\kappa=0.3$. This function is
compared with the ordinary exponential and with a pure power
law.}\label{fig1}
\end{figure}

In fig. 2,  the function $\exp_{_{\{{\scriptstyle \kappa}\}}}(-x)$
for some different values of $\kappa$ is plotted. We note that
when $\kappa \rightarrow 0$ the $\kappa$-exponential approaches
the ordinary one.
\begin{figure}[ht]
\centerline{
\includegraphics[width=.8\columnwidth,angle=-90]{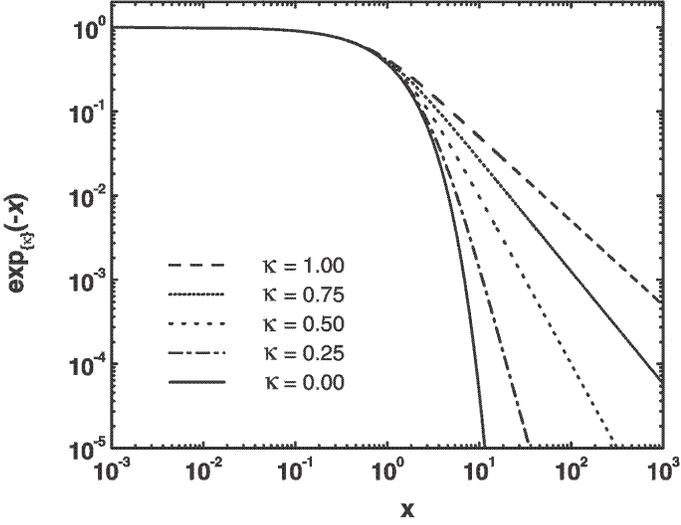}}
\caption{Plot of the function $\exp_{_{\{{\scriptstyle
\kappa}\}}}(-x)$ versus $x$ for some different values of $\kappa$.
The case $\kappa =0$ corresponds to the ordinary
exponential.}\label{fig2}
\end{figure}
The Taylor expansion of $\ln_{_{\{{\scriptstyle \kappa}\}}}(1+x)$
converges if $-1<x\leq 1$ and assumes the form
\begin{equation}
\ln_{_{\{{\scriptstyle \kappa}\}}}(1+x) = \sum_{n=1}^{\infty}
b_n(\kappa)\,(-1)^{n-1} \,\frac{x^n}{n} \ \ , \label{V24}
\end{equation}
(\cite{MATHG}, p. 25), with $b_1(\kappa)= 1$ while for $n>1$,
$b_n(\kappa)$ is given by
\begin{eqnarray}
b_{n}(\kappa)&&\,\,=\frac{1}{2}\Big(1-\kappa\Big)\Big(1-\frac{\kappa}{2}\Big)...
\Big(1-\frac{\kappa}{n-1}\Big) \nonumber
\\ &&\,\,+\,\frac{1}{2}\Big(1+\kappa\Big)\Big(1+\frac{\kappa}{2}\Big)...
\Big(1+\frac{\kappa}{n-1}\Big) \label{V25} \ \ .
\end{eqnarray}
It results $b_n(0)=1$ and $b_n(-\kappa)=b_n(\kappa)$. The first
terms of the expansion are
\begin{equation}
\ln_{_{\{{\scriptstyle \kappa}\}}}(1+x) = x - \frac{x^2}{2} +
\left(1+\frac{\kappa^2}{2}\right)\frac{x^3}{3} - ... \ \ .
\label{V26}
\end{equation}

Another expansion involving the $\kappa$-exponential $(x^2\leq 1)$
is the following
\begin{eqnarray}
\exp_{_{\{{\scriptstyle \kappa}\}}}(x)= \exp
\Big(\sum_{n=0}^{\infty} d_n\,\kappa^{2n}x^{2n+1}\Big) \ \ ,
\label{V27}
\end{eqnarray}
(\cite{MATHG}, p. 58), being
\begin{eqnarray}
d_n=\frac{(\!-1)^n\,(2n)!}{(2n\!+\!1)\,2^{2n}\,(n!)^2} \ \ .
\label{V28}
\end{eqnarray}
Exploiting this expansion, we can write the $\kappa$-exponential
as an infinite product of standard exponentials
\begin{eqnarray}
\exp_{_{\{{\scriptstyle \kappa}\}}}(x)=\prod_{n=0}^{\infty} \exp
\Big(d_n\,\kappa^{2n}x^{2n+1}\Big) \ \ . \label{V29}
\end{eqnarray}

On the other hand the $\kappa$-exponential can be viewed as a
continuous linear combination of an infinity of standard
exponentials. Namely for ${\rm Re}\,s \geq 0$ it results
(\cite{MATHG}, p. 1108)
\begin{equation}
\exp_{_{\{{\scriptstyle \kappa}\}}}(-s) =
\int_0^{\infty}\frac{1}{\kappa x}\,\,J_{_{\scriptstyle
\!1/\kappa}}\Big(\frac{x}{\kappa}\Big) \exp \,(-sx)\, dx \ \ .
\label{V30}
\end{equation}

The following two integrals can be useful

\begin{eqnarray}
&&\int_{\, 0}^{\infty}\!x^{r-1} \,\exp_{_{\{{\scriptstyle
\kappa}\}}}\!(-x)\,dx \nonumber \\
&&=\frac{\big[1+(r-2)|\kappa|\,\big]\,|2\kappa|^{-r}}{\big[1-(r\!-\!1)\,
|\kappa|\big]^2-\kappa^2}\,\,\,
\frac{\Gamma\left(\frac{1}{|2\kappa|}-\frac{r}{2}
\right)}{\Gamma\left(\frac{1}{|2\kappa|}+\frac{r}{2}
\right)}\,\,\Gamma\left(r\right)\,\,\,\,\,\,\,\,\, \label{V31}
\end{eqnarray}

\begin{eqnarray}
&&\int_{\, 0}^{1}\left(\ln_{_{\{{\scriptstyle
\kappa}\}}}\!\frac{1}{x}\right)^{r-1}\,dx \nonumber \\
&&=\frac{|2\kappa|^{1-r}}{1+(r-1)|\kappa|}\,\,\,\frac{\Gamma\left(\frac{1}{|2\kappa|}-\frac{r-1}{2}
\right)}{\Gamma\left(\frac{1}{|2\kappa|}+\frac{r-1}{2} \right)}
\,\,\Gamma\left(r\right). \label{V32}
\end{eqnarray}

We conclude the present section focusing our attention to another
interesting property of $\kappa$-logarithm. We consider the
following eigenvector equation
\begin{eqnarray}
{\cal D}_{\kappa}(x)\, L(x)=l_{\kappa}\,L(x)\ \ , \label{V33}
\end{eqnarray}
and examine the case of the eigenvector $L(x)=n^x$. It is trivial
to verify that when ${\cal D}_{0}(x)=d/dx$ we obtain that the
eigenvalue $l_0$ of this operator is the standard logarithm
$l_{0}=\ln n$. We pose now the question if it is possible to
determine the operator ${\cal D}_{\kappa}(x)$ associated to the
same eigenvector $L(x)=n^x$ and having as eigenvalue the
$\kappa$-logarithm, namely
\begin{equation}
l_{\kappa}=\ln_{_{\{{\scriptstyle \kappa}\}}}n \ \ . \label{V34}
\end{equation}
We obtain that this operator is the finite difference operation
\begin{eqnarray}
{\cal D}_{\kappa}(x)\, L(x)=\frac{L(x+\kappa)-L(x-\kappa)}{2
\kappa} \ \ , \label{V35}
\end{eqnarray}
which reduces to the standard derivative as the increment
$2\kappa$ of the independent variable approaches to zero.

One can find many other elegant and useful mathematical
properties for the $\kappa$-functions which obviously we can not
report here.

\sect{Relativistic kinetics}

In the present section we treat the statistical system, considered
previously in stationary conditions, within a relativistic and
kinetic framework. The new relativistic kinetics here presented,
in the limit $c\rightarrow \infty$, reduces to the classical
kinetics considered in \cite{PHA01}.

By using the standard notation of the relativistic theory we
denote with $x=x^{\nu}=(ct,\mbox{\boldmath $x$})$ the four-vector
position and with $p=p^{\nu}=(p^{0},\mbox{\boldmath $p$})$ the
four-vector momentum, being $p^{0}=\sqrt{\mbox{\boldmath $p$}^2 +
m^2c^2}$ and employ the metric $g^{\mu\nu}=diag\,(1,-1,-1,-1)$
\cite{DEGROOT}.

Let us consider the following relativistic kinetic equation
\begin{eqnarray}
p^{\,\nu}\partial_{\nu}f-m F^{\nu}\frac{\partial f}{\partial
p^{\,\nu}}= \int
\frac{d^3p'}{{p'}^{0}}\frac{d^3p_1}{p_1^{\,0}}\frac{d^3p'_1}{{p'}_{\!\!1}^{0}}
\,\,G \,\,&& \nonumber \\
\times
\left[\,a\,(f'\otimes\mbox{\raisebox{-2mm}{\hspace{-3.3mm}$\scriptstyle
\kappa$}} \hspace{1mm}f'_1)-
a\,(f\otimes\mbox{\raisebox{-2mm}{\hspace{-3.3mm}$\scriptstyle
\kappa$}} \hspace{1mm}f_1)\,\right],&& \label{VI1}
\end{eqnarray}
where the distribution $f$ is a function of the four-vectors $x$
and $p$, namely $f=f(x,p)$. We note that the left hand side of
Eq. (\ref{VI1}) is the same of the standard relativistic
Boltzmann equation but the collision integral in the right hand
side results to be more complicated, containing  the deformed
product $\otimes\mbox{\raisebox{-2mm}{\hspace{-2mm}$\scriptstyle
\kappa$}} \hspace{1mm}$ and the arbitrary function $a(f)$ which
we suppose to be positive and increasing. The factor $G$ is the
transition rate which depends only on the nature of the two body
particle interaction.

The above equation in the case $\kappa=0$ and $a(f)=f$ reduces to
the already known relativistic Boltzmann equation describing the
standard relativistic kinetics \cite{DEGROOT}. Clearly in the
case $\kappa \neq 0$ and $a(f)\neq f$, the above equation
describes a new relativistic kinetics, radically different from
the standard one.

We anticipate that this new relativistic kinetics, which we will
consider here, defines a statistics resulting to be independent
on the particular form of the function $a(f)$.

{\it Steady states:} We consider now the steady states of Eq.
(\ref{VI1}) for which the collision integral becomes equal to
zero.  Then we have
\begin{eqnarray}
f\otimes\mbox{\raisebox{-2mm}{\hspace{-3.3mm}$\scriptstyle
\kappa$}}\hspace{1mm}f_1=
f'\otimes\mbox{\raisebox{-2mm}{\hspace{-4.8mm}
$\scriptstyle\kappa$}} \hspace{1mm}f'_1 \ \ , \label{VI2}
\end{eqnarray}
and after taking into account the property (\ref{II61}) of the
$\kappa$-logarithm, we obtain
\begin{equation}
\ln_{_{\{{\scriptstyle \kappa}\}}}f+ \ln_{_{\{{\scriptstyle
\kappa}\}}}f_{1} =\ln_{_{\{{\scriptstyle
\kappa}\}}}f^\prime+\ln_{_{\{{\scriptstyle
\kappa}\}}}f^\prime_{1} \ . \ \ \label{VI3}
\end{equation}
This last equation represents a conservation law and then we can
conclude that $\ln_{_{\{{\scriptstyle \kappa}\}}}f$ is a
summational invariant; in the most general case it is a linear
combination of the microscopic relativistic invariants, namely a
constant and the four-vector momentum. In ref. \cite{DEGROOT} it
is shown that in presence of external electromagnetic fields the
more general microscopic relativistic invariant has the form
$\left(p^{\nu}+q A^{\nu}\!/c \right)\,U_{\nu}+$ constant, being
$U_{\nu}$  the hydrodynamic four-vector velocity with
$U^{\nu}U_{\nu}=c^2$. Then we can pose
\begin{equation}
\ln_{_{\{{\scriptstyle \kappa}\}}}f=-\frac{\left(p^{\nu}+q
A^{\nu}\!/c \right)\,U_{\nu}-mc^2-\mu }{\lambda\,k_{_{B}}T}\, \ .
\label{VI4}
\end{equation}
Consequently we obtain the following stationary distribution
\begin{equation}
f= \exp_{_{\{{\scriptstyle \kappa}\}}}\bigg(-\frac{\left(p^{\nu}+q
A^{\nu}\!/c \right)\,U_{\nu}-mc^2-\mu }{\lambda\,k_{_{B}}T}\bigg)
\ . \label{VI5}
\end{equation}
In the case $\kappa=0$ this distribution reduces to the already
known relativistic distribution \cite{DEGROOT}.

The above equilibrium distribution, in the global rest frame where
$U^{\nu}=(c,0,0,0)$ and in absence of external forces
($A^{\nu}=0$), simplifies as
\begin{equation}
f=\exp_{_{\{{\scriptstyle \kappa}\}}}\!\left(-\,
\frac{E-\mu}{\lambda\,k_{_{B}}T}\right) \ \ , \label{VI6}
\end{equation}
and assumes the same form of the distribution (\ref{III1}).

We remark that for $E-\mu\gg \lambda\, k_{_{B}}T $ this
distribution presents a power law behavior, namely
\begin{equation}
f \approx \left(\frac{E_*}{E}\right)^{1/\kappa}  \ \ , \label{VI7}
\end{equation}
being $E_*=k_{_{B}}T\,\sqrt{1-\kappa^2}\,/2\kappa$.

In order to introduce explicitly the dependence on the velocity
variable in the distribution (\ref{VI6}), we consider the
expression of the relativistic kinetic energy $E=E(v)$
\begin{equation}
E=\sqrt{m^2c^4+{\mathbf p}^2c^2}-mc^2 \ \ , \label{VI8}
\end{equation}
with ${\mathbf p}=m\gamma(v){\mathbf v}$ and the Lorentz factor
given by
\begin{equation}
\gamma(v)=\frac{1}{\sqrt{1-v^2/c^2}}  \ \ . \label{VI9}
\end{equation}
After defining $\eta=\mu/mc^2$ we write Eq. (\ref{VI6}) as follows
\begin{equation}
f = \exp_{_{\{{\scriptstyle
\kappa}\}}}\!\left(-\frac{m\,c^2}{\,\lambda\,k_{_{B}}T}\,\big[\gamma(v)-1
-\eta \big] \right) \ \ . \label{VI10}
\end{equation}
Note that in the region $v \ll c$ this distribution assumes the
form
\begin{equation}
f \approx \exp_{_{\{{\scriptstyle
\kappa}\}}}\!\left(-\frac{\frac{1}{2}mv^2-\mu}{\,\lambda\,
k_{_{B}}T}\, \right) \ \ , \label{VI11}
\end{equation}
which, after setting $\lambda\, k_{_{B}}T=1/\beta$, coincides with
the non relativistic statistical distribution, proposed in ref.
\cite{PHA01}.

{\it H-theorem:} In the standard relativistic kinetics it is well
known from the H-theorem that the production of entropy is  never
negative and in equilibrium conditions there is no entropy
production. In the following we will demonstrate the H-theorem
for the system governed by the kinetic equation (\ref{VI1}). For
simplicity of the notation, hereafter we omit the letter $\kappa$
in the symbol of $\kappa$-entropy. We define the four-vector
entropy $S=S^{\nu}=(S^{0},\mbox{\boldmath $S$})$, in terms of the
distribution $n=\alpha f$, as follows
\begin{equation}
S^{\nu}= - k_{_B}\int \frac{d^3p}{p^{0}}\,p^{\nu}\, n
\,\ln_{_{\{{\scriptstyle \kappa}\}}}\!n \ , \label{VI12}
\end{equation}
and note that $S^{0}$ coincides with the $\kappa$-entropy defined
previously through Eq. (\ref{IV20}) while $\mbox{\boldmath $S$}$
is the entropy flow. If we take into account the relation
(\ref{III9}), the above four-vector entropy can be written in
terms of the distribution $f$ as
\begin{equation}
S^{\nu}= - k_{_B}\lambda \,\alpha \int
\frac{d^3p}{p^{0}}\,p^{\nu}\,\int df \,\ln_{_{\{{\scriptstyle
\kappa}\}}}\! f \ . \label{VI13}
\end{equation}
It is trivial to verify that the entropy production
$\partial_{\nu}S^{\nu}$ can be calculated starting from the
definition of $S^{\nu}$ and the evolution equation (\ref{VI1}),
obtaining
\begin{eqnarray}
\partial_{\nu}S^{\nu}= &&- k_{_B}\lambda\,\alpha \int
\frac{d^3p}{p^{0}}\,(\ln_{_{\{{\scriptstyle
\kappa}\}}}\!f)\,p^{\nu}\,\partial_{\nu}f \nonumber \\
=&&- k_{_B}\lambda \,\alpha \int
\frac{d^3p'}{{p'}^{0}}\frac{d^3p_1}{p_1^{\,0}}\frac{d^3p'_1}{{p'}_{\!\!1}^{0}}
\frac{d^3p}{p^{0}}\,\,G \nonumber \\
&& \times
\left[\,a\,(f'\otimes\mbox{\raisebox{-2mm}{\hspace{-3.3mm}$\scriptstyle
\kappa$}} \hspace{1mm}f'_1)-
a\,(f\otimes\mbox{\raisebox{-2mm}{\hspace{-3.3mm}$\scriptstyle
\kappa$}} \hspace{1mm}f_1)\,\right] \, \ln_{_{\{{\scriptstyle \kappa}\}}}\!f \nonumber \\
&&- k_{_B}\lambda\,\alpha \,m \int
\frac{d^3p}{p^{0}}\,(\ln_{_{\{{\scriptstyle
\kappa}\}}}\!f)\,F^{\nu}\frac{\partial f}{\partial p^{\,\nu}}. \
\ \  \label{VI14}
\end{eqnarray}
Since the Lorentz force $F^{\nu}$ has the properties
$p^{\nu}F_{\nu}=0$ and $\partial F^{\nu}/\partial p^{\nu}=0$ the
last term in the above equation involving $F^{\nu}$ is equal to
zero \cite{DEGROOT}, namely
\begin{eqnarray}
\partial_{\nu}S^{\nu}=\!\!\!\!\! &&- k_{_B}\lambda \,\alpha \int
\frac{d^3p'}{{p'}^{0}}\frac{d^3p_1}{p_1^{\,0}}\frac{d^3p'_1}{{p'}_{\!\!1}^{0}}
\frac{d^3p}{p^{0}}\,\,G \nonumber \\
&& \times
\left[\,a\,(f'\otimes\mbox{\raisebox{-2mm}{\hspace{-3.3mm}$\scriptstyle
\kappa$}} \hspace{1mm}f'_1)-
a\,(f\otimes\mbox{\raisebox{-2mm}{\hspace{-3.3mm}$\scriptstyle
\kappa$}} \hspace{1mm}f_1)\,\right] \, \ln_{_{\{{\scriptstyle
\kappa}\}}}\!f . \ \ \  \label{VI15}
\end{eqnarray}
Given the particular symmetry of the integral in Eq. (\ref{VI15})
we can write the entropy production as follows
\begin{eqnarray}
\partial_{\nu}S^{\nu}=\!\!\!\!\!&&-\frac{1}{4}\, k_{_B} \lambda\,\alpha \int
\frac{d^3p'}{{p'}^{0}}\frac{d^3p_1}{p_1^{\,0}}\frac{d^3p'_1}{{p'}_{\!\!1}^{0}}
\frac{d^3p}{p^{0}}\,\,G \nonumber \\
&&\times
\left[\,a\,(f'\otimes\mbox{\raisebox{-2mm}{\hspace{-3.3mm}$\scriptstyle
\kappa$}} \hspace{1mm}f'_1)-
a\,(f\otimes\mbox{\raisebox{-2mm}{\hspace{-3.3mm}$\scriptstyle
\kappa$}} \hspace{1mm}f_1)\,\right] \nonumber \\
&&\times \, [\ln_{_{\{{\scriptstyle \kappa}\}}}\!f +
\ln_{_{\{{\scriptstyle \kappa}\}}}\!f_1-\ln_{_{\{{\scriptstyle
\kappa}\}}}\!f' - \ln_{_{\{{\scriptstyle \kappa}\}}}\!f'_1 ]. \ \
\ \ \  \label{VI16}
\end{eqnarray}
Finally, we set this equation in the form
\begin{eqnarray}
\partial_{\nu}S^{\nu}=\!\!\!\!\!&&\frac{1}{4}\, k_{_B}\lambda\,\alpha \int
\frac{d^3p'}{{p'}^{0}}\frac{d^3p_1}{p_1^{\,0}}\frac{d^3p'_1}{{p'}_{\!\!1}^{0}}
\frac{d^3p}{p^{0}}\,\,G \nonumber \\
&&\times
\left[\,a\,(f'\otimes\mbox{\raisebox{-2mm}{\hspace{-3.3mm}$\scriptstyle
\kappa$}} \hspace{1mm}f'_1)-
a\,(f\otimes\mbox{\raisebox{-2mm}{\hspace{-3.3mm}$\scriptstyle
\kappa$}} \hspace{1mm}f_1)\,\right] \nonumber \\
&&\times \left[\,\ln_{_{\{{\scriptstyle \kappa}\}}}\!
(f'\otimes\mbox{\raisebox{-2mm}{\hspace{-3.3mm}$\scriptstyle
\kappa$}} \hspace{1mm}f'_1)- \ln_{_{\{{\scriptstyle
\kappa}\}}}\!(f\otimes\mbox{\raisebox{-2mm}{\hspace{-3.3mm}$\scriptstyle
\kappa$}} \hspace{1mm}f_1)\right]. \label{VI17}
\end{eqnarray}
After imposing that $a(h)$ is an increasing function, it results
$[a(h_1)-a(h_2)]\,[\ln_{_{\{{\scriptstyle
\kappa}\}}}\!(h_1)-\ln_{_{\{{\scriptstyle \kappa}\}}}\!(h_2)]\geq
0$  $\forall h_1,h_2$  and then we can conclude that
\begin{equation}
\partial_{\nu}S^{\nu}\geq 0 \ \ . \label{VI18}
\end{equation}
This last relation is the local formulation of the relativistic
H-theorem which represents the second law of the thermodynamics
for the system governed by the evolution equation (\ref{VI1}).

Concerning the arbitrary positive and increasing function $a(f)$
appearing in the collision integral of the evolution equation, we
note that, if we suppose that obeys to the following condition
\begin{eqnarray}
a \, (f\otimes\mbox{\raisebox{-2mm}{\hspace{-3.3mm}$\scriptstyle
\kappa$}} \hspace{1mm}f_1)=a \,(f) \, a(f_1) \ \ , \label{VI19}
\end{eqnarray}
we recover the expression
\begin{eqnarray}
a(f)=\exp \big(\ln_{_{\{{\scriptstyle \kappa}\}}}f\,\big) \ \ ,
\label{VI20}
\end{eqnarray}
proposed in ref. \cite{PHA01}.

\sect{Physical meaning of the $\kappa$-deformation}

In the present section we will show that the deformation
introduced by the parameter $\kappa$ emerges naturally within the
Einstein's special relativity, so that one can see the
$\kappa$-deformation as a purely relativistic effect.

Let us consider in the one-dimensional frame $\cal S$ two
identical particles of rest mass $m$. We suppose that the first
particle moves toward right with velocity $v_1$ while the second
particle moves toward left with velocity $v_2$. The relativistic
momenta of the particles are given by $p_1=p\,(v_1)$ and
$p_2=p\,(v_2)$ respectively, being
\begin{equation}
p\,(v)= \frac{m\,v}{\sqrt{1-v^2/c^2}} \ \ . \label{VII1}
\end{equation}

We consider now the same particles in a new frame $\cal S\,'$
which moves at constant speed $v_2$ toward left with respect to
the frame $\cal S$. In this new frame the particles have
velocities given by $v\,'_1=v_1\oplus^c v_2$ and $v\,'_2=0$
respectively, being
\begin{equation}
v_1\oplus^c v_2=\frac{v_1+v_2}{1+v_1v_2/c^2} \label{VII2}
\end{equation}
the well known relativistic additivity law for the velocities. In
the same frame $\cal S\,'$ the particle relativistic momenta are
given by $p\,'_1=p\,(v\,'_1)$ and $p\,'_2=0$, respectively. Up to
now, we have simply recalled some well known concepts of the
special relativity \cite{LI}.

Let us pose the following question: if it is possible and how to
obtain the value of the relativistic momentum $p\,'_1$ starting
directly from the values of the momenta $p_1$ and $p_2$ in the
frame $\cal S$. The answer to this apparently innocent question
is affirmative. One, after straightforward calculations (see the
theorem in this section), arrives to the following surprising
result
\begin{equation}
p\,'_1=p\,(v_1)\oplus\!\!\!\!\!^{^{\scriptstyle \kappa}}\,\,
p\,(v_2) \ \ \ ; \ \ \ \kappa=\frac{1}{m c} \ \ . \label{VII3}
\end{equation}
In words, the relativistic momentum $p\,'_1$ of the first
particle in the rest frame of the second particle is the
$\kappa$-deformed sum, with $\kappa=1/mc$, of the momenta $p_1$
and $p_2$ of the particles in the frame $\cal S$.

Unexpectedly we discover that the $\kappa$-sum is the additivity
law for the relativistic momenta. Eq.(\ref{VII3}) which we write
in the form
\begin{equation}
p\,(v_1)\oplus\!\!\!\!\!^{^{\scriptstyle \kappa}}\,\, p\,(v_2)=
p\,(v_1\oplus^c v_2) \ \ \ ; \ \ \ \kappa=\frac{1}{m c} \ \ ,
\label{VII4}
\end{equation}
says that the $\kappa$-deformed sum and the relativistic sum of
the velocities are intimately related and reduce both, to the
standard sum as the velocity $c$ approaches to infinity. The
deformations in both the cases are relativistic effects and are
originated from the fact that $c$ has a finite value.
Eq.(\ref{VII4}) follows as a particular case from the following
theorem:

{\it Theorem:} Let
\begin{equation}
p_i\,(v_i)= \frac{m_i\,v_i}{\sqrt{1-v_i^2/c^2}} \ \ , \label{VII5}
\end{equation}
be the relativistic momenta of two particles $(i=1,2)$ of rest
mass $m_1$ and $m_2$ which move in the one dimensional frame
$\cal S$ with speed $v_1$ and and $v_2$ respectively. If we
indicate with $\oplus\!\!\!\!^{^{\scriptstyle \kappa}}$ the
$\kappa$-sum defined through Eq. (\ref{V11}) and with $\oplus^c$
the velocity relativistic additivity law defined through Eq.
(\ref{VII2}), it results that
\begin{equation}
\frac{p_1\,(v_1)}{m_1}\oplus\!\!\!\!\!^{^{\scriptstyle
\kappa}}\,\, \frac{p_2\,(v_2)}{m_2}= \frac{p_i\,(v_1\oplus^c
v_2)}{m_i} \ \ ; \ \ \kappa=\frac{1}{c} \ \ \ . \label{VII6}
\end{equation}

{\it Proof:} We start by using the definition of the $\kappa$-sum,
subsequently we use the explicit form of the relativistic momentum
and finally we use the definition of the velocity relativistic
additivity law

\begin{eqnarray}
&& \frac{p_1\,(v_1)}{m_1}\oplus\!\!\!\!\!^{^{\scriptstyle
\kappa}}\,\, \frac{p_2\,(v_2)}{m_2} \nonumber \\
&& = \frac{p_1\,(v_1)}{m_1} \sqrt{1+
\left[\frac{p_2\,(v_2)}{m_2\,c}\right]^2}+ \frac{p_2\,(v_2)}{m_2}
\sqrt{1+
\left[\frac{p_1\,(v_1)}{m_1\,c}\right ]^2} \nonumber \\
&& = \frac{v_1}{\sqrt{1-\left(v_1/c \right)^2}} \,\sqrt{1+
\frac{(v_2/c)^2}{1-\left(v_2/c\right)^2}}
\nonumber \\
&& + \frac{v_2}{\sqrt{1-\left(v_2/c \right)^2}} \,\sqrt{1+
\frac{(v_1/c)^2}{1-\left(v_1/c\right)^2}}
\nonumber \\
&& = \frac{v_1+v_2}{ \sqrt{\left[1-(v_1/c)^2 \right]
\left[1-(v_2/c)^2 \right]}} \nonumber \\ && = \left(v_1 \oplus^c
v_2\right) \, \frac{1+v_1 v_2/c^2}{ \sqrt{\left[1-(v_1/c)^2
\right] \left[1-(v_2/c)^2 \right]}} \nonumber \\ && = \frac{v_1
\oplus^c
v_2}{\sqrt{1-\displaystyle{\frac{1}{c^2}\left(\frac{v_1+v_2}
{1+v_1v_2/c^2}\right)^2}}}
 \nonumber \\ && =
\frac{v_1 \oplus^c v_2}{ \sqrt{ 1-\displaystyle{\frac{\left(v_1
\oplus^c v_2\right)^2}{c^2}} } }
 \nonumber \\ && =
 \frac{p_i\,(v_1\oplus^c
v_2)}{m_i} \ \ \ . \label{VII6a}
\end{eqnarray}

 Trivially from Eq. (\ref{VII6}) one obtains
Eq.(\ref{VII4}) as particular case, when $m_1=m_2=m$. Note that
the parameter $\kappa$ has different values in these two
equations because the summed quantities in the two cases are
different.

We can easily explain the meaning of the deformed derivative. We
indicate with $p$ the relativistic momentum in the frame $\cal S$,
and with $dG/dp$ the derivative with respect to $p$ of the Lorentz
invariant scalar $G$. The same quantities in the frame $\cal S\,'$
are indicated with $p\,'$ and $dG/dp\,'$, respectively.  It is
trivial to verify that
\begin{equation}
\frac{d\, G}{d \, p_{_{\{{\scriptstyle \kappa}\}}}} = \frac{d\,
G}{d \, p\,'} \ \ , \label{VII5}
\end{equation}
and then we can conclude that the $\kappa$-deformed derivative can
be viewed as a standard derivative in an appropriate frame.

In the next section we will consider, in the framework of the
special relativity, the $\kappa$-statistics of $N$-identical
particles, where $\kappa$ is a dimensionless parameter and we
will determine its value. To do so, it is more convenient to
write Eq. (\ref{VII4}) in the form
\begin{equation}
\frac{p\,(v_1)}{\kappa\, m\, c}\oplus\!\!\!\!\!^{^{\scriptstyle
\kappa}}\,\,\frac{p\,(v_2)}{\kappa\, m\, c}=\frac{p\,(v_1\oplus^c
v_2)}{\kappa\, m\, c} \ \ . \label{VII7}
\end{equation}
which holds for any $\kappa$.

\sect{Determination of the parameter $\kappa$}

In the present section we calculate the value of the parameter
$\kappa$ which, due to the symmetry $\kappa \leftrightarrow
-\kappa$ of the theory, we consider positive, namely
$0\leq\kappa<1$. Clearly if we start from the $\kappa$-statistical
distributions $f_1$ and $f_2$ describing two independent
statistical systems, we can construct the distribution
$f_1\otimes\mbox{\raisebox{-2mm}{\hspace{-2.8mm}$\scriptstyle
\kappa$}} \hspace{1mm}f_2$ which describes a new composite system.
This system in the case $\kappa=0$ reduces to the one described
through the distribution $f_1\,f_2$. In the following we will
assume that {\it the distribution $f_1\,f_2$ describes a state
also in the case $\kappa\neq 0$}. Obviously this state is
different from the one described by
$f_1\otimes\mbox{\raisebox{-2mm}{\hspace{-2.8mm}$\scriptstyle
\kappa$}} \hspace{1mm}f_2$. As we will see, this simple but
meaningful hypothesis is sufficient to determine the value of
$\kappa$ in the case of relativistic systems.

Taking into account the form of the distributions $f_1$ and $f_2$
given by Eq. (\ref{VI6}) (for simplicity we pose $\mu=0$), and the
property (\ref{II30}) of the $\kappa$-exponential one can write
immediately
\begin{eqnarray}
\exp_{_{\{{\scriptstyle \kappa}\}}}\!\left(-\,
\frac{E_1}{\lambda\,k_{_{B}}T}\right)\exp_{_{\{{\scriptstyle
\kappa}\}}}\!\left(-\,
\frac{E_2}{\lambda\,k_{_{B}}T}\right)\nonumber
\\ = \exp_{_{\{{\scriptstyle \kappa}\}}}\!\left(-
\frac{E_1}{\lambda\,k_{_{B}}T}\oplus\!\!\!\!\!^{^{\scriptstyle
\kappa}}\,\, \frac{E_2}{\lambda\,k_{_{B}}T}\right) \ \ .
\label{VIII1}
\end{eqnarray}
with $E_i=E(v_i)$. After some simple algebra we rewrite Eq.
(\ref{VIII1}) as follows
\begin{eqnarray}
\exp_{_{\{{\scriptstyle \kappa}\}}}\!\left(-\,
\frac{E_1}{\lambda\,k_{_{B}}T}\right)\exp_{_{\{{\scriptstyle
\kappa}\}}}\!\left(-\,
\frac{E_2}{\lambda\,k_{_{B}}T}\right)\nonumber
\\ = \exp_{_{\{{\scriptstyle \kappa}\}}}\!\left(-
\frac{E_3}{\lambda\,k_{_{B}}T}\right) \ \ , \label{VIII2}
\end{eqnarray}
with $E_3=E_1\sqrt{1+E_2^2/E_0^2}+E_2\sqrt{1+E_1^2/E_0^2}$ and
\begin{eqnarray}
E_0=\frac{\lambda\,k_{_{B}}T}{\kappa} \ \ . \label{VIII3}
\end{eqnarray}
We assume now that the $\kappa$-exponential in the right hand side
of the Eq. (\ref{VIII2}) has the same structure of the one given
by Eq. (\ref{VI6}). Clearly we must impose that $E_0$ be
exclusively expressed in terms of non statistical parameters. In
the following we will show that $E_0=mc^2$. To do this we exploit
Eq. (\ref{VII7}) obtained within the special relativity. Starting
from this equation and taking into account the property
(\ref{II30}) of the $\kappa$-exponential we obtain
\begin{eqnarray}
\exp_{_{\{{\scriptstyle \kappa}\}}}\!\left(-\,
\frac{p\,(v_1)}{\kappa\, m\, c}\right)\exp_{_{\{{\scriptstyle
\kappa}\}}}\!\left(-\, \frac{p\,(v_2)}{\kappa\, m\,
c}\right)\nonumber
\\ = \exp_{_{\{{\scriptstyle \kappa}\}}}\!\left(-
\frac{p\,(v_1\oplus^c v_2)}{\kappa\, m\, c} \right) \ \ .
\label{VIII4}
\end{eqnarray}

Recall that we wish to calculate the parameter $\kappa$ which has
a value that does not depend on the particle energy. Then,
without loosing generality, we can consider Eq. (\ref{VIII4}) in
the ultrarelativistic region $(v \rightarrow c)$ where it results
$p\,(v) \approx E(v)/c\,\,$:
\begin{eqnarray}
\exp_{_{\{{\scriptstyle \kappa}\}}}\!\left(-\,
\frac{E\,(v_1)}{\kappa\, m\, c^2}\right)\exp_{_{\{{\scriptstyle
\kappa}\}}}\!\left(-\, \frac{E\,(v_2)}{\kappa\, m\,
c^2}\right)\nonumber
\\ = \exp_{_{\{{\scriptstyle \kappa}\}}}\!\left(-
\frac{E\,(v_1\oplus^c v_2)}{\kappa\, m\, c^2} \right) \ \ .
\label{VIII5}
\end{eqnarray}
After comparing Eq.s (\ref{VIII2}) and (\ref{VIII5}), one obtains
the relation
\begin{equation}
\kappa \, m \, c^2=\lambda\,k_{_{B}}T \ \ , \label{VIII6}
\end{equation}
which is the same as the one given by Eq. (\ref{VIII3}), only if
we impose that $E_0=mc^2$. Eq.(\ref{VIII6}) can be also written as
\begin{equation}
k_{_{B}}T=mc^2\frac{\kappa}{\sqrt{1-\kappa^2}} \ \ , \label{VIII7}
\end{equation}
and results to be formally similar $(k_{_{B}}T/c \leftrightarrow
p\ , \ \ \kappa \leftrightarrow v/c)$ to the relation defining
the relativistic momentum given by Eq. (\ref{VII1}). We can
extract finally the value of $\kappa$ obtaining
\begin{equation}
\frac{1}{\kappa^2}=1+\displaystyle{\left(\frac{m\,c^2}
{k_{_{B}}T}\right)^2} \ \ . \label{VIII8}
\end{equation}
It is important to emphasize that this expression of the
parameter $\kappa$ holds only under the above mentioned hypothesis
and imposes that $|\kappa| <1$. This condition on the value of
$\kappa$ coincides with the one expressed by Eq. (\ref{IV15}) and
obtained in a completely different way. We have $\kappa=0$ only
if $T=0$ or if $c=\infty$. The limiting case $\kappa=1$ is
obtained if $T=\infty$ or if $mc^2=0$.

At this point one can write the distribution (\ref{VI6}) in the
form
\begin{equation}
f=\exp_{_{\{{\scriptstyle \kappa}\}}}\!\left(-\,\frac{1}{\kappa}\,
\frac{E-\mu}{mc^2}\right) \ \ . \label{VIII9}
\end{equation}
Note that the statistical information of the system, namely the
temperature is hidden exclusively in the parameter $\kappa$. When
$E \rightarrow \infty$ the distribution (\ref{VIII9}) shows a
power law asymptotic behavior
\begin{equation}
f\approx \left(\frac{mc^2}{2E}\right)^{1/\kappa} \ \ .
\label{VIII10}
\end{equation}

The distribution (\ref{VIII9}) viewed as a function of the
velocity becomes
\begin{equation}
f=\exp_{_{\{{\scriptstyle \kappa}\}}}\!\left(-
\frac{\gamma(v)-1-\eta}{\kappa}\right) \ \ ; \ \ v<c \ \ .
\label{VIII11}
\end{equation}
Concerning its derivative one obtains
\begin{equation}
\frac{d\,f}{d\,v}=-\frac{v}{\kappa\,c^2}\frac{\gamma^3}
{\sqrt{1+(\gamma-1-\eta)^2}} \, f \ \ , \label{VIII12}
\end{equation}
and then for $v\rightarrow c$ it results
\begin{equation}
\frac{d\,f}{d\,v}\approx-\frac{1}{c\,\kappa\,2^{1+1/2\kappa}
}\left(1-\frac{v}{c}\,\right)^ {-1+1/2\kappa} \ \ . \label{VIII13}
\end{equation}
Then for $\kappa<1/2$ one has both $f=0$ and $df/dv=0$ in $v=c$.
For $\kappa>1/2$ results $f=0$ and $df/dv=-\infty$ in $v=c$.
Finally for $\kappa=1/2$ results $f=0$ and $df/dv=-1/2c$ in $v=c$.

In the non relativistic region for which $v \ll c$ we have
\begin{equation}
f \approx \exp_{_{\{{\scriptstyle
\kappa}\}}}\!\left(-\frac{\frac{1}{2}mv^2-\mu}{\,\kappa\, m c^2}\,
\right) \ \ , \label{VIII14}
\end{equation}
while in the limit $c \rightarrow \infty$ one recovers the
standard Maxwellian distribution
\begin{equation} f_{_M} = \exp
\left(-\frac{\frac{1}{2}mv^2-\mu}{\,k_{_{B}}T}\, \right) \ \ .
\label{VIII15}
 \label{N28}
\end{equation}

The explicit form of the distribution (\ref{VIII11}) when
$\mu=-mc^2$ simplifies as follows
\begin{equation}
f=\left(\frac{\sqrt{1-v^2/c^2}}{1+\sqrt{2-v^2/c^2}}\right)^{1/\kappa}\
\ \ ; \ \ v<c \ \ , \label{VIII16}
\end{equation}
and in the limit $c \rightarrow \infty$ becomes
\begin{equation}
f_{_M} = \exp \left(-\frac{mv^2}{2\,k_{_{B}}T}\, \right) \ \ .
\label{VIII17}
\end{equation}

In fig. 3, the distribution function given by Eq. (\ref{VIII16})
(after normalization) versus $v/c$ for different values of
$\kappa$ and then for different values of $mc^2/k_{_{B}}T$
according to Eq. (\ref{VIII8}), is plotted.
\begin{figure}[ht]
\centerline{
\includegraphics[width=.8\columnwidth,angle=-90]{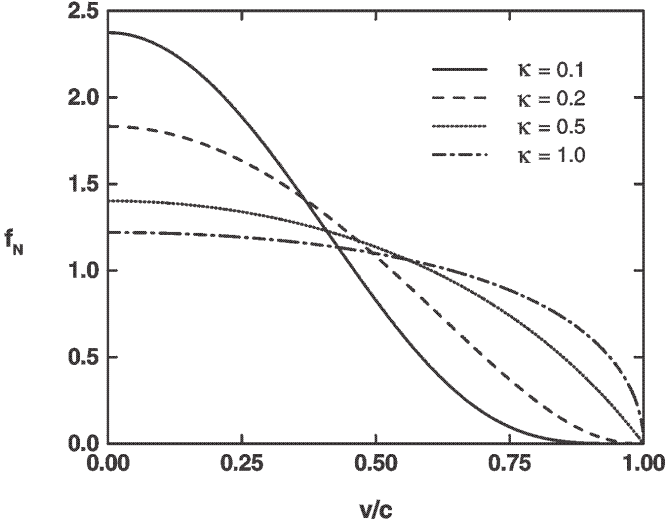}}
\caption{Plot of the distribution function (after normalization)
given by Eq. (\ref{VIII16}) versus $v/c$ for some different
values of $\kappa$.}\label{fig3}
\end{figure}

\sect{Experimental Evidences}

Since long time it is known that the cosmic rays spectrum, which
extends over $13$ decades in energy, from a few hundred of MeV
($10^8$ eV) to a few hundred of EeV ($10^{20}$ eV) and spans $33$
decades in particle flux, from $10^{4}$ to $10^{-29}$ $[m^2\,
sr\, s\, GeV]^{-1}$, is not exponential and then it violates the
Boltzmann equilibrium statistical distribution $\propto \exp
(-E/k_{_{B}}T)$ \cite{COSRAYG,CRT,COSRAYD}. Approximately this
spectrum follows a power law $E^{-a}$ and the spectral index $a$
is near 2.7 below $5\times 10^{15}$ eV, near 3.1 above $5\times
10^{15}$ eV and again near 2.7 above $3\times 10^{18}$ eV. On the
other hand it is known that the particles composing the cosmic
rays are essentially the normal nuclei as in the standard cosmic
abundances of matter. Then the cosmic rays can be viewed as an
equivalent statistical system of identical relativistic particles
with masses near the mass of the proton ($938$ MeV).

These above characteristics (relativistic particles with a very
large extension both for their flux and energy) yield the cosmic
rays spectrum an ideal physical system for a preliminary test of
the correctness and predictability of the theory here proposed.

We consider the statistical distribution $f(E)$ given by Eq.
(\ref{VI6}) or (\ref{VIII9}). The particle flux
$\Phi(E)\propto\,p^2 f(E)$ can be calculated trivially if we take
into account the relativistic expression linking $E$ and $p$
obtaining
\begin{equation}
\Phi(E)=A\,\left[\left(\frac{E}{mc^2}+1\right)^{2}-1\right]\,
\exp_{_{\{{\scriptstyle
\kappa}\}}}\!\big(\!\!-\beta\,\,(E-\mu)\,\big) \ \ . \label{IX1}
\end{equation}
Note that this particle flux, in agreement with the observational
data, decays following the power law
\begin{equation}
\Phi(E)\propto E^{\,-a} \ \ , \label{IX2}
\end{equation}
with
\begin{equation}
a=\sqrt{1+\displaystyle{\left(\frac{m\,c^2}
{k_{_{B}}T}\right)^2}}\, -\, 2 \ \ . \label{IX3}
\end{equation}

Analogously the particle flux obtained starting from the
Boltzmann-Gibbs statistical distribution is given by

\begin{equation}
\Phi_0(E)=A_0\,\left[\left(\frac{E}{mc^2}+1\right)^{2}-1\right]\,
\exp\left(-\frac{E}{k_{_{B}}T}\right) \ \ . \label{IX4}
\end{equation}

We use these two theoretical distributions of particle flux to fit
the cosmic rays data reported in ref. \cite{COSRAYD}. In fig. 4,
we show the observed data together with the theoretical curves
$\Phi(E)$ (solid line) and $\Phi_0(E)$ (dotted line). The curve
$\Phi(E)$ corresponding to $A=10^{5}$ $[m^2\, sr\, s\, GeV]^{-1}$,
$mc^2=938\,MeV$, $\mu=-375\,MeV$, and $\kappa=0.2165$ provides a
high quality agreement with the observed data. This agreement
over so many decades is quite remarkable. From the value of
$\kappa$ and $mc^2$ and adopting Eq. (\ref{VIII7}) we obtain that
$k_{_{B}}T=208\,MeV$.
\begin{figure}[ht]
\centerline{
\includegraphics[width=.8\columnwidth,angle=0]{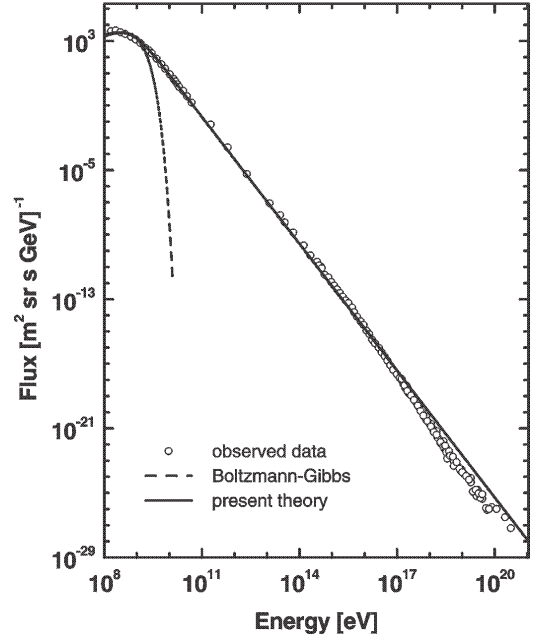}}
\caption{Plot of the cosmic rays flux versus energy. The solid
line is the curve obtained within the present theory and is given
by Eq. (\ref{IX1}) with $A=10^{5}$ $[m^2\, sr\, s\, GeV]^{-1}$,
$mc^2=938\,MeV$, $\mu=-375\,MeV$, $\kappa=0.2165$ and
$k_{_{B}}T=208\,MeV$. The dotted line represents the theoretical
curve obtained within the standard Boltzman-Gibbs statistics
given by Eq. (\ref{IX4})  with $A_0=1.3 \cdot 10^4$ $[m^2\, sr\,
s\, GeV]^{-1}$, $mc^2=938\,MeV$ and $k_{_{B}}T=208\,MeV$. The
observational data are collected by S.P. Swordy
\cite{COSRAYD}.}\label{fig4}
\end{figure}

In the same figure the curve $\Phi_0(E)$ (with $A_0=1.3 \cdot
10^4$ $[m^2\, sr\, s\, GeV]^{-1}$, $mc^2=938\,MeV$ and
$k_{_{B}}T=208\,MeV$) which decays exponentially and can not fit
the observed data violating the Boltzmann-Gibbs statistics, is
reported .

A short remark must be made at this point. Clearly, the power law
asymptotic behavior of the spectrum $\Phi(E)$ is imposed by the
$\kappa$-exponential whose origin is the $\kappa$-sum. But the
$\kappa$-sum emerges naturally within the special relativity as
the composition law of the relativistic momenta. Then we can
conclude that the power law asymptotic behavior of the cosmic
rays flux is simply the signature of the particle relativistic
nature.

It is widely known that the Boltzmann-Gibbs distribution $\exp
(-E/k_{_{B}}T)$ originally proposed to describe a classical
particle gas in thermodynamic equilibrium can be adopted to
describe an enormous amount of phenomena in nature. On the other
hand the power law tails have been observed experimentally in
several fields of science. Some times in particular fields, this
power law has a name (e.g. Pareto law in econophysics,
Gutenberg-Richter law in seismology etc). Furthermore the power
law tail is preceded by an exponential region and between the two
regions exists a third intermediate region. It is worth remarking
that the $\kappa$-exponential defines a distribution which can
describe simultaneously the three above regions (see fig.1) and
then is particularly suitable to describe the above mentioned
phenomena.

As a working example we analyze the experimental data reported in
ref. \cite{OLE2} related to the rain events in meteorology. In
fig. 5 is plotted the number density $N$[events/(year mm)] of rain
events versus the event size $M$[mm] on a double logarithmic
scale. We note that the data have a large extension (the abscissa
spans 7 decades and the ordinate 5) and remark that its behavior
is typical of a class of experimental data which we find in
several areas of science. In order to fit the experimental data
we adopt the distribution
\begin{equation}
N=A \, \exp_{_{\{{\scriptstyle
\kappa}\}}}\!\big(\!\!-\beta\,M\,\big) \ \ , \label{IX5}
\end{equation}
and, as one can see in fig. 5, a remarkable agreement is obtained
with $A=8\cdot10^4$ [events/(year mm)], $\beta=75$ [mm$^{-1}$] and
$\kappa=0.7$.
\begin{figure}[ht]
\centerline{
\includegraphics[width=.8\columnwidth,angle=-90]{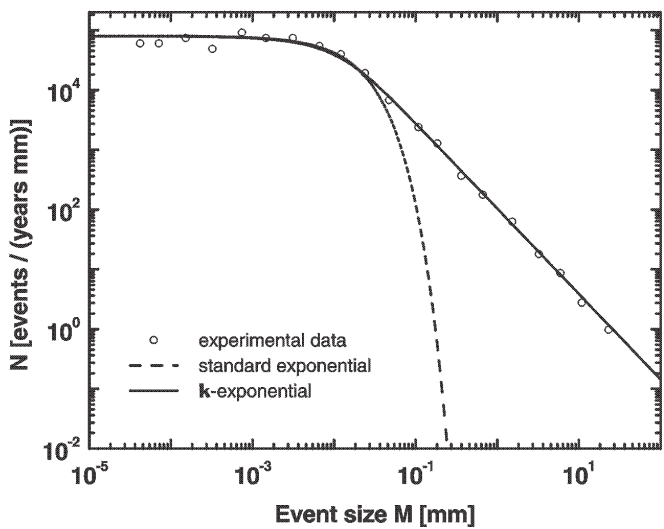}}
\caption{Plot of the number density $N$[events/(year mm)] of rain
events versus the event size $M$[mm]. The solid line is the curve
obtained within the present theory and is given by Eq.
(\ref{IX5}) with $A=8\cdot10^4$ [events/(year mm)], $\beta=75$
[mm$^{-1}$] and $\kappa=0.7$. The dotted line corresponds to the
ordinary exponential function ($\kappa=0$). The experimental data
are from ref. \cite{OLE2}.}\label{fig5}
\end{figure}

Clearly, one can hunt other mechanisms, different from the
relativistic one, leading to $\kappa$-statistics. Beside the
important problem of the agreement between the theoretical curve
and the observational data one does not neglect the
epistemological problem concerning the structure of the theory
which must be able both to explain the origin and to determine
the value of any parameter appearing within the theory.

\sect{Conclusions}

We summarize briefly the results obtained in the present effort.
We have shown that beside the Boltzmann-Shannon entropy, the
quantity
\begin{eqnarray}
S_{\kappa}=- \,k_{_B}\!\sum_i n_i\, \ln_{_{\{{\scriptstyle
\kappa}\}}}\!n_i   \ \ , \nonumber
\end{eqnarray}
with $\ln_{_{\{{\scriptstyle
\kappa}\}}}\!x=(x^{\kappa}-x^{-\kappa})/2\kappa$ and
$-1<\kappa<1$, is the only existing entropy, simultaneously
concave, additive and extensive. Starting from this entropy it is
possible to construct a generalized statistical mechanics (and
thermodynamics) having the same mathematical and epistemological
structure of the Boltzmann-Gibbs one, which is recovered when the
deformation parameter $\kappa$ approaches to zero. Within this
generalized statistics the distribution function assumes the form
\begin{eqnarray}
 n_i= \alpha \,\exp_{_{\{{\scriptstyle \kappa}\}}}(-\beta
\,(\,E_i-\mu\,)\,) \ \ , \nonumber
\end{eqnarray}
with $\exp_{_{\{{\scriptstyle \kappa}\}}}(x)=
\left(\sqrt{1+\kappa^2x^2}+\kappa x\right)^{1/\kappa}$ while the
constants $\alpha$ and $\beta$ are given by
$\alpha=[(1-\kappa)/(1+\kappa)]^{1/2\kappa}$,
$1/\beta=\sqrt{1-\kappa^2}\,\,k_{_{B}}\!T$. The chemical
potential $\mu$ can be fixed by the normalization condition. This
distribution has a bulk very close to the exponential one while
its tail decays following a power law $n_i\propto
E_i^{-1/\kappa}$.

The origin of the deformation has its roots in the Einstein
special relativity and the relativistic statistical systems are
governed by a kinetics obeying the H-theorem.

We have shown that, within the special relativity, it is possible
to determine the value of $\kappa$, obtaining in this case
\begin{eqnarray}
\frac{1}{\kappa^2}=1+\displaystyle{\left(\frac{m\,c^2}
{k_{_{B}}T}\right)^2} \ \ , \nonumber
\end{eqnarray}
so that the relativistic statistical meechanics does not contain
free parameters.

The theory can describe observational data in many fields. In
particular we find a high quality agreement in analyzing the
spectrum of the cosmic rays which violates manifestly the
Boltzmann-Gibbs statistics. This is an important test for the
theory because the cosmic rays are relativistic particles and
their spectrum has a very large extension ($13$ decades in energy
and $33$ decades in flux).

\vfill
\eject

\begin{thebibliography}{99}

\bibitem{PHA01} G. Kaniadakis, Physica A {\bf 296}, 405 (2001).
\bibitem{PLA01} G. Kaniadakis, Phys. Lett. A {\bf 288}, 283 (2001).
\bibitem{PHA02} G. Kaniadakis and A.M. Scarfone,
Physica A {\bf 305}, 69 (2002).
\bibitem{NAU} J. Naudts, Physica A, in press (2002), [cond-mat/0203489].
\bibitem{ABED} S. Abe and Y. Okamoto, {\it Nonextensive Statistical
Mechanics and its Applications}, Lectures Notes in Physics {\bf
560}, Spring-Verlag Berlin (2001).
\bibitem{EDI02} G. Kaniadakis, M. Lissia and A. Rapisarda Eds.,
Special Issue of Physica A {\bf 305},(2002).
\bibitem{OLE1} O. Peters, C. Hertlein, and K.Christensen, Phys. Rev. Lett.
{\bf 88}, 018701-1 (2002)
\bibitem{OLE2} O. Peters, and K.Christensen, Phys. Rev. E {\bf 66}, 036120 (2002).
\bibitem{ECON} H.E. Stanley, L.A.N. Amaral, X. Gabaix,
P. Gopikrishnan, V. Plerou, Physica A {\bf 299}, 1 (2001).
\bibitem{TLOG} C. Tsallis, Quimica Nova {\bf 17}, 468 (1994).
\bibitem{ABE} S. Abe, Phys. Lett. A {\bf 224}, 326 (1997).
\bibitem{BORO} E.P. Borges and I. Roditi, Phys. Lett. A {\bf 246}, 399 (1998).
\bibitem{JAY1} E.T. Jaynes, Phys. Rev. {\bf 106}, 620 (1957); {\bf 108}, 171 (1957).
\bibitem{HACHA} J. Harvda and F. Charvat, Kybernetica {\bf 3}, 30
(1967).
\bibitem{NEXT} C. Tsallis, J. Stat. Phys. {\bf 53}, 479 (1988)
\bibitem{VASYL} V.M. Vasyliunas, J. Geophys. Res. {\bf 73}, 2839 (1968).
\bibitem{MATHG} I.S. Gradshteyn  and I.M. Ryzhik, {\it Table of integrals, series, and
Product}, Academic Press London (2000).
\bibitem{TUS} J.A. Tuszynaki, J.L. Rubin, J. Meyer, and M. Kibler,
Phys. Lett. A {\bf 175}, 173 (1993).
\bibitem{UBR} M.R. Ubriaco, Phys. Rev. E {\bf 57}, 179 (1998).
\bibitem{DEGROOT} S.R. de Groot, W.A. van Leeuwen, and Ch.G. van Weert, {\it Relativistic Kinetic Theory: Principle and Applications},
North-Holland, Amsterdam (1980).
\bibitem{LI} R.L. Libboff,  {\it Kinetic Theory}, Prentice-Hall
International, Englewoord Cliffs, NJ, (1990).
\bibitem{COSRAYG} P.L. Biermann, and G. Sigl, {\it Physics and Astrophysics of
Ultra-Hight-Energy Cosmic Rays}, Lectures Notes in Physics {\bf
576}, Spring-Verlag Berlin (2001), [astro-ph/0202425].
\bibitem{CRT} C. Tsallis, J.C. Anjos, and E.P. Borges,
arXiv:astro-ph/0203258.
\bibitem{COSRAYD} S.P. Swordy, Cosmic Rays Data, Chicago
University, http://astroparticle.uchicago.edu/announce.htm.
\end{thebibliography}
\end{document}